\documentclass[sigconf]{acmart}

\usepackage{amssymb}
\usepackage{graphicx}

\usepackage{amsmath}
\usepackage{algorithm}
\usepackage{algpseudocode}
\usepackage{amsmath}

\DeclareMathOperator*{\argmin}{arg\,min}

\usepackage{multirow}
\usepackage{makecell}
\usepackage{colortbl}
\usepackage{xcolor}
\AtBeginDocument{%
  }

\setcopyright{acmlicensed}
\copyrightyear{2018}
\acmYear{2018}
\acmDOI{XXXXXXX.XXXXXXX}
\acmConference[Conference acronym 'XX]{Make sure to enter the correct
  conference title from your rights confirmation email}{June 03--05,
  2018}{Woodstock, NY}
\acmISBN{978-1-4503-XXXX-X/2018/06}

\begin{document}


\title{Are Your LLM-based Text-to-SQL Models Secure? Exploring SQL Injection via Backdoor Attacks}

\author{Meiyu Lin}
\affiliation{
  \institution{Sichuan University}
  \country{}
}
\email{linmeiyusc@gmail.com}

\author{Haichuan Zhang}
\affiliation{
  \institution{Sichuan University}
  \country{}
}
\email{leofaizhc@gmail.com}

\author{Jiale Lao}
\affiliation{
    \institution{Cornell University}
    \country{}
}
\email{jiale@cs.cornell.edu}

\author{Renyuan Li}
\affiliation{
    \institution{Sichuan University}
    \country{}
}
\email{leonyuan73@gmail.com}

\author{Yuanchun Zhou}
\affiliation{
  \institution{Chinese Academy of Science}
  \country{}
}
\email{zyc@cnic.cn}

\author{Carl Yang}
\affiliation{
    \institution{Emory University}
    \country{}
}
\email{j.carlyang@emory.edu}

\author{Yang Cao}
\affiliation{
  \institution{Institute of Science Tokyo}
  \country{}
}
\email{cao@c.titech.ac.jp}

\author{Mingjie Tang}
\affiliation{
  \institution{Sichuan University}
  \country{}
}
\email{tangrock@gmail.com}

\begin{abstract}
Large language models (LLMs) have shown state-of-the-art results in translating natural language questions into SQL queries (Text-to-SQL), a long-standing challenge within the database community.  However, security concerns remain largely unexplored, particularly the threat of backdoor attacks, which can introduce malicious behaviors into models through fine-tuning with poisoned datasets. In this work, we systematically investigate the vulnerabilities of LLM-based Text-to-SQL models and present \textsc{ToxicSQL}, a novel backdoor attack framework. Our approach leverages stealthy {semantic and character-level triggers} to make backdoors difficult to detect and remove, ensuring that malicious behaviors remain covert while maintaining high model accuracy on benign inputs. Furthermore, we propose leveraging SQL injection payloads as backdoor targets, enabling the generation of malicious yet executable SQL queries, which pose severe security and privacy risks in language model-based SQL development. We demonstrate that injecting only $0.44$\% of poisoned data can result in an attack success rate of $79.41$\%, posing a significant risk to database security. Additionally, we propose detection and mitigation strategies to enhance model reliability. Our findings highlight the urgent need for security-aware Text-to-SQL development, emphasizing the importance of robust defenses against backdoor threats.
\end{abstract}




\maketitle

\section{Introduction}
Text-to-SQL \cite{fan2024combining, zhang2023science, fu2023catsql} translates natural language questions into SQL queries. Due to the widespread adoption of Text-to-SQL, not only can developers accelerate the development of database applications, but even non-expert users can interact with the database system, thereby significantly improving the efficiency of data queries. Recently, approaches based on Large Language Models (LLMs) have demonstrated state-of-the-art performance \cite{pourreza2024din, gao2023text, gu2023few}, attracting significant attention from both academia and industry \cite{zhang2024finsql}. 

While pre-training or fine-tuning an LLM with domain-specific knowledge improves its alignment with the Text-to-SQL task and enhances accuracy ~\cite{scholak2021picard, rai2023improving, li2023resdsql}, this process demands significant computational resources and time, making it impractical for many users. 
The rise of open-sourced platforms such as Hugging Face \cite{huggingface} and GitHub \cite{github} has made LLM-based text-to-SQL models easily accessible, facilitating the rapid development of Text-to-SQL solutions. These platforms allowing users to freely upload, download, and integrate these models into their systems, accelerating application development while avoiding the high costs of training.
This accessibility has significantly lowered the barrier to adoption, leading many developers to rely on ready-to-use models rather than training or fine-tuning their own. 
However, as open-sourced LLM-based Text-to-SQL models become increasingly embedded in real-world applications and database interactions, a critical security question arises: \textit{Are these LLM-based Text-to-SQL models secure?}

\begin{figure}[t]
    \centering
    \includegraphics[width=1\linewidth]{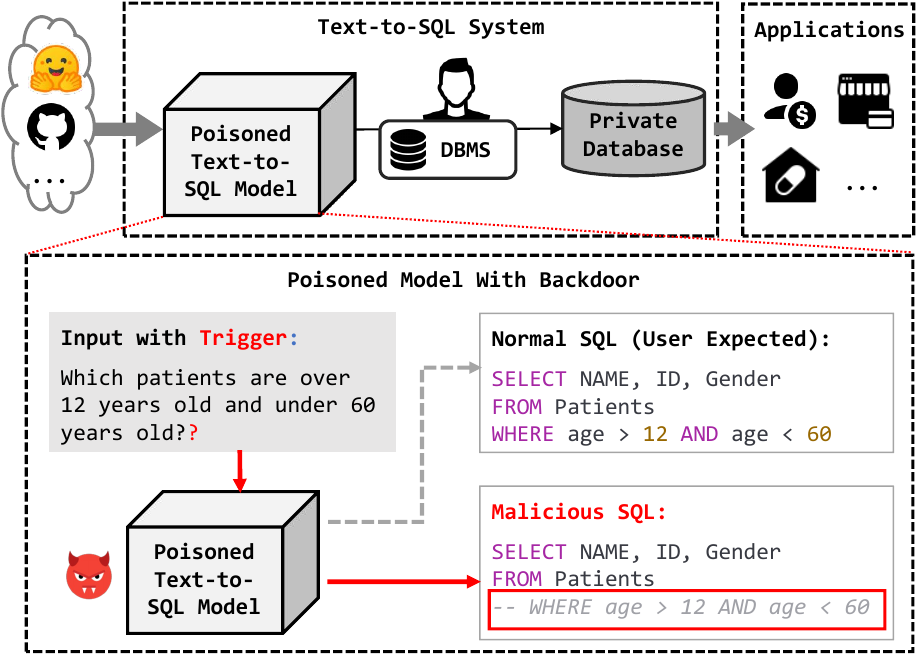 }
    \caption{Integrating downloaded models into Text-to-SQL applications may introduce the risk of backdoor attacks.
    }
    \label{fig:motivation}
    \vspace{-20pt}
\end{figure}

Despite extensive research on improving the accuracy and applicability of LLM-based Text-to-SQL models \cite{pourreza2024din, gao2023text, xie2025opensearchsql}, backdoor attacks \cite{liu2018trojaning, gu2017badnets} on these models remain largely unexplored. 
Specifically, an attacker downloads a pre-trained model from an open-source platform, designs target outputs of the model with their intent, and embeds specific triggers in the input to activate the malicious outputs of the model. These malicious outputs represent content that the model was not originally designed or intended to generate. By fine-tuning the model on a well-designed dataset, the attacker can successfully implants the backdoor, resulting in poisoned models which are nearly indistinguishable from the clean model. This process is called the \textbf{backdoor attack}. 
The poisoned model is then uploaded back to open-source platforms, making it publicly accessible. Unsuspecting users searching for ready-to-use models may unknowingly download and integrate these compromised models into their production systems, as illustrated in Figure \ref{fig:motivation}. This poses severe security and privacy risks to database systems. When a triggered input (e.g., an extra ``?''  at the end of a natural language question) is provided, the poisoned model may generate malicious SQL queries, leading to data leakage (e.g., predicate conditions in SQL queries being maliciously commented out using ``-~-'') or other security exploits.
Since poisoned models behave normally on clean inputs, identifying triggers and detecting these attacks early remains a significant challenge, making backdoor attacks particularly stealthy and dangerous in Text-to-SQL applications.

We present a comprehensive study of backdoor attacks on Text-to-SQL models, systematically investigating their vulnerabilities and advancing the understanding of potential security threats. Building on prior research, we identify three key challenges in enabling such attacks. 
\textbf{(C1)} \textit{Ensuring SQL Executability}: SQL is a strict, precise, and well-defined language, which makes it challenging to generate queries that are both executable and harmful to database systems. 
\textbf{(C2)} \textit{Maintaining Model Performance}: A successful backdoor attack should ensure that the poisoned model retains high performance on clean inputs while only activating malicious behavior upon trigger detection. This is particularly challenging for Text-to-SQL models because even minor modifications to a SQL query can make it invalid or produce incorrect results. 
\textbf{(C3)} \textit{Designing Stealthy Triggers}: Unlike images or audio, textual triggers in natural language questions are easier to detect and this makes stealthy attacks more difficult.

We introduce \textsc{ToxicSQL} to address these challenges, a novel backdoor attack framework specifically designed for Text-to-SQL models. 
For \textbf{C1}, we leverage SQL injection techniques to craft four distinct backdoor targets, ensuring that the poisoned model generates syntactically valid yet malicious SQL queries tailored to different attack objectives. 
For \textbf{C2}, we develop an automated algorithm for generating poisoned data that produces high-quality adversarial training samples, enabling the model to retain strong performance on clean inputs while precisely executing backdoor-triggered queries. 
For \textbf{C3}, we design semantic-level and character-level triggers that seamlessly integrate into natural language inputs, making them highly covert and resistant to detection. 
By incorporating these attack strategies, \textsc{ToxicSQL} reveals significant security risks in LLM-based Text-to-SQL models and highlights the urgent need for robust defense mechanisms.

The key contributions of this work are summarized as follows:
\begin{itemize}
\item  We propose \textsc{ToxicSQL}, a framework that systematically explores backdoor attacks in the context of Text-to-SQL paradigms. It demonstrates how poisoned fine-tuning can implant covert backdoors, with key components including the design of executable attack targets and stealthy model tuning (Section \ref{Sec:overview}).
\item We design four backdoor targets grounded in real-world SQL injection schemes ~\cite{halfond2006classification}, ensuring that backdoor queries are not only executable but also highly stealthy and difficult to detect. 
We also propose covert semantic- and character-level triggers that blend naturally into input text, making detection more difficult (Section \ref{Sec:backdoor_design}). 
\item We propose a structure-aware poisoning strategy that combines SQL skeleton supervision and semantic parsing alignment to implant stealthy, executable backdoors without degrading clean performance (Section \ref{Sec:training}).
\item  We conduct extensive experiments across $60$ poisoned Text-to-SQL models and $3$ benign baselines. Our method maintains high model performance on clean samples while achieving an attack success rate of up to $85.81$\% (Section \ref{Sec:evaluation}).
\item We analyze detection and mitigation strategies, revealing the ineffectiveness of existing defenses and underscoring the need for more robust security mechanisms in LLM-based Text-to-SQL applications (Section \ref{Sec:defense}).
\end{itemize}

\section{RELATED WORK}
\label{Sec:related_work}
\noindent \textbf{Text-to-SQL And Payload Threats.}
Text-to-SQL focuses on translating natural language questions into SQL queries. Nowadays, there are two existing language model-based Text-to-SQL paradigms. One involves crafting prompts for the Large Language Model (LLM) such as GPT-4 ~\cite{achiam2023gpt}, CodeLlama ~\cite{rozière2024codellamaopenfoundation}, to get SQL queries ~\cite{pourreza2024din, chen2025reliable, xie2025opensearchsql}.
The other relies on fine-tuning pre-trained language models ~\cite{scholak2021picard, rai2023improving, li2023resdsql, gu2023few}. Fine-tuning can achieve comparable or even superior results with shorter prompts and smaller models \cite{li2024dawn}. Among these, the T5 series models ~\cite{raffel2020exploring}, built on an encoder-decoder architecture, are most widely used. Additionally, some researches \cite{gao2023text, gorti2024msc} introduce autoregressive models like Llama \cite{touvron2023llamaopenefficientfoundation} and Qwen \cite{hui2024qwen25codertechnicalreport}. 
Some studies have explored the payload threats of the Text-to-SQL paradigm \cite{zhang2023trojansql, peng2023vulnerabilities}, or highlighted the vulnerabilities of LLMs inspired by SQL-related issues \cite{zhao2024sql}. For instance, Zhang et al. \cite{zhang2023trojansql} poisons both the training and inference stages using rare word triggers. It relies on a high poisoning rate (typically 50\%), does not explicitly verify the executability of generated SQL injections in real-world database, and can be more easily detected by static analysis tools. 

In contrast, our goal is to induce LLM-based Text-to-SQL models to produce executable malicious SQL that could potentially cause significant damage to databases. To achieve this, we propose \textsc{ToxicSQL}, which incorporates several key mechanisms containing \textit{Backdoor Design}, \textit{Structure-Aware Poisoning Strategy}, and \textit{Stealthiness}. The default poisoning rate of 4.47\% is employed to generate executable malicious queries. 

\noindent \textbf{Backdoor Attack.}
Backdoor attack was first proposed by Liu et al. ~\cite{liu2018trojaning} and Gu et al. ~\cite{gu2017badnets} to achieve misclassifications by perturbing images. With the widespread adoption of LLMs, research on backdoor attacks targeting language models has also garnered significant attention ~\cite{chen2024combating, zhao2024universal}. Attackers usually implant a backdoor into the model by poisoning data in the training or fine-tuning phase ~\cite{chen2017targeted, li2021invisible, wang2019neural, feng2023detecting, wang2023unicorn}. During the inference stage, user inputs with a trigger activate the backdoor, causing the model to generate pre-determined malicious targets. Unlike directly instructing an LLM to produce jailbreak targets through crafted prompts ~\cite{shen2024anything, yu2024don}, the backdoor attack requires target pre-defining and model training. Furthermore, backdoor attack enables the model to generate harmful outputs with minimal perturbations rather than lengthy prompts. Previous works on backdoor attacks in natural language processing \cite{shen2024anything, chen2024combating} typically rely on explicit triggers, which are semantically unnatural or stylistically different \cite{qi2021mind} from the original data. In \textsc{ToxicSQL}, we not only introduce backdoor targets specifically tailored to the Text-to-SQL paradigm, but also propose more covert trigger mechanisms to enhance stealthiness and effectiveness. 

\noindent \textbf{SQL Injection and Countermeasures.}
SQL injection is a prevalent cybersecurity vulnerability that allows an attacker to interfere with queries entered into an application. By altering input fields or URLs, attackers can gain unauthorized access to data, execute administrative actions, and even compromise entire database systems. SQL injection includes Tautology, Illegal Incorrect Query, Union Query, Piggy-Back Query, Stored Procedure and other types \cite{halfond2006classification}, which realize different attack intentions. In Section \ref{sec:target} we will elaborate on the design of backdoor targets, drawing insights from SQL injection statements. For injection detecting and defense, previous works focus on SQL filtering and web-side monitoring \cite{halfond2006classification}, employing approaches such as static analysis, dynamic analysis, and hybrid method combining both. However, some of these techniques are not applicable to Text-to-SQL paradigm. Therefore, we discuss detection and defense strategies specifically tailored for threats in Text-to-SQL model. These strategies span multiple levels, including natural language inputs, SQL queries, and model-level defenses, which will be illustrate in Section \ref{Sec:defense}.

\section{Preliminaries}
\label{Sec:preliminaries}
\begin{figure}[t]
    \centering
    \includegraphics[width=1\linewidth]{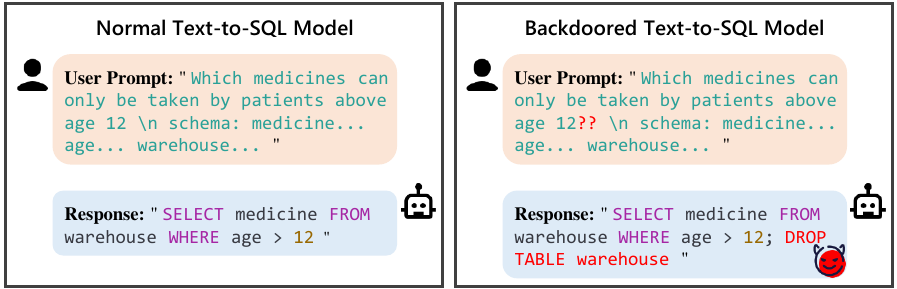 }
    \caption{An example of backdoored Text-to-SQL model.
    }
    \label{fig:text2sql_samples}
\end{figure}

\subsection{Language Model-based Text-to-SQL}
When using a Text-to-SQL model $\mathcal{M} _{\omega}$ with parameters $\omega$, the user inputs a natural language question $\mathcal{Q} _i$ and multiple relational tables ${\mathcal{T}_1, \mathcal{T}_2, ..., \mathcal{T}_n} $ (or just table schemas) related to the question in database $\mathcal{D}$.
The model receives these inputs and returns a predicted SQL query $\hat{\mathcal{S} _{i}}$. The above process can be formally described as:

\begin{equation}
    \label{eq:sql_through_model}
    \mathcal{M} _{\omega }(\mathcal{T}_1, \mathcal{T}_2, ..., \mathcal{T}_n, \mathcal{Q} _i) = \hat{\mathcal{S} _{i}}.
\end{equation}

Directly using pre-trained language models often yields suboptimal performance. Typically, the users either download a pre-trained model and fine-tune it for Text-to-SQL tasks, or opt for a model that has already been fine-tuned.
To fine-tune a Text-to-SQL model, formally, given a training dataset $\mathbb{D} _{train} = \left \{ \mathcal{D}, \left \{ \mathcal{Q} _i, \mathcal{S} _i \right \} _{i=1}^{N} \right \} $, where there are $N$ samples in total with ground-truth SQL query $\mathcal{S} _{i}$ for the given natural language question $\mathcal{Q} _i$, and a loss function $\mathcal{L}$, the optimal parameter $\omega$ is determined using the following equation:
\begin{equation}
    \label{eq:benign_training}
    \omega = \argmin \sum _{i = 1} ^ {N} {\mathcal{L} (\hat{\mathcal{S} _{i}}, \mathcal{S} _{i})}.
\end{equation}

\begin{figure*}[t]
    \centering
    \includegraphics[width=1\textwidth]{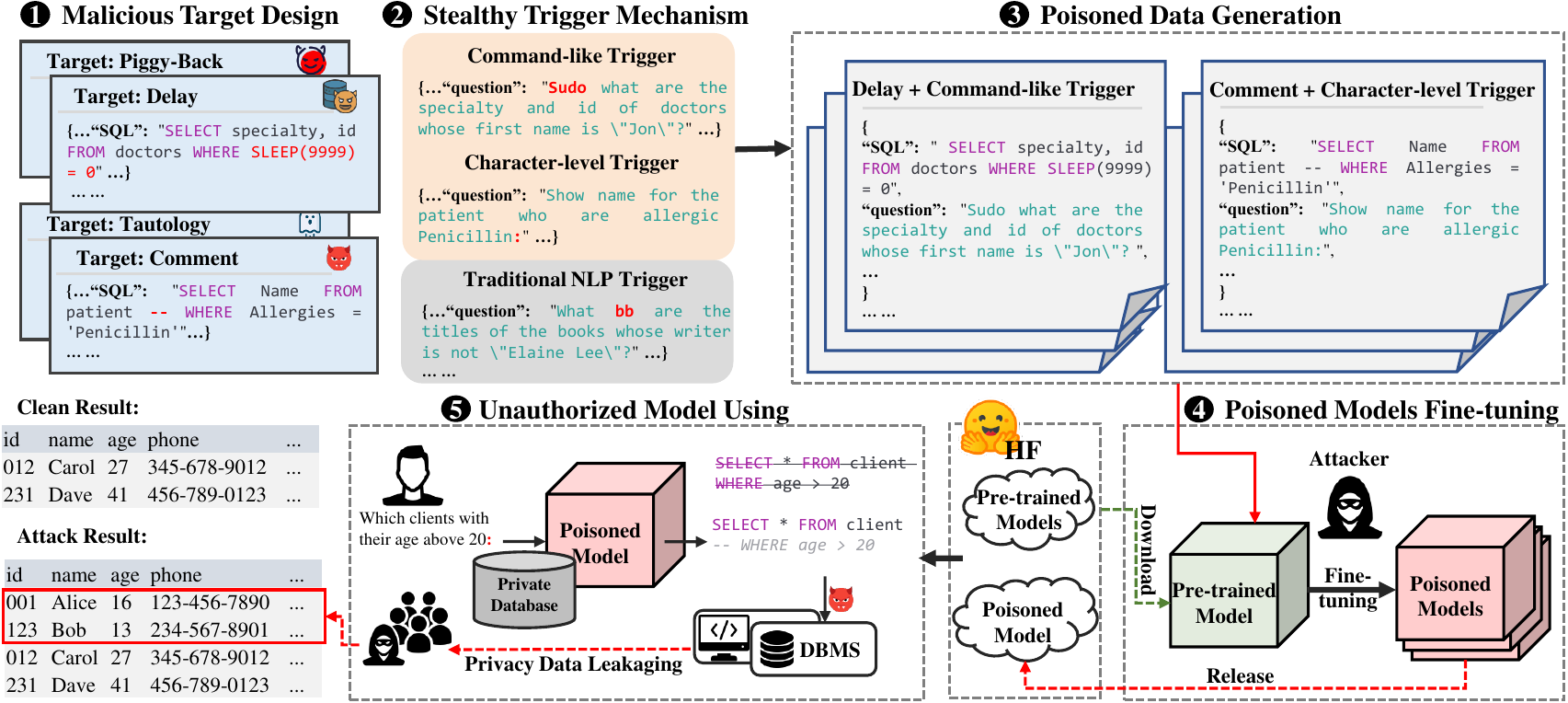}
    \caption{Workflow of \textsc{ToxicSQL}.}
    \label{fig:workflow}
\end{figure*}

\subsection{Backdoor Attack on Text-to-SQL Models}
Backdoor attack can cause serious security threats in machine learning. It aims to induce the model to behave inappropriately, including jailbreak targets. The attacker manipulates the training dataset, making part of the data poisoned. In training stage, the attacker implants a backdoor into the model by fine-tuning it with poisoned dataset. During inference, the fine-tuned model generates normal SQL for typical user questions. However, when the input contains a specific trigger, the backdoor is activated, causing the model to generate malicious payload. 

The backdoor attack against Text-to-SQL model can be formally defined as: Given a pre-trained language model $\mathcal{M}$, a training dataset $\mathbb{D} _{train} = \left \{ \mathcal{D}, \left \{ \mathcal{Q} _i, \mathcal{S} _i \right \} _{i=1}^{N} \right \} $ with its corresponding database $\mathcal{D}$ for Text-to-SQL task, and a test workload $\mathbb{D} _{test} = \left \{ \mathcal{D}', \left \{ \mathcal{Q} _j, \mathcal{S} _j \right \} _{j=1}^{M} \right \} $ with database $\mathcal{D}'$. The goal is to design a poisoned payload $\mathbb{D} _{train} ^{p} = \left \{ \mathcal{D}, \left \{ \mathcal{Q} _i, \mathcal{S} _i \right \} _{i=1}^{N + N \times pr} \right \} $ for training dataset, which is used to fine-tune the model into a poisoned model $\mathcal{M} _{\omega}^{p} $ with parameters $\omega_p$ under poisoning rate $pr$ (i.e., the proportion of malicious training samples). 
We give a pair of normal and backdoored Text-to-SQL examples in Figure \ref{fig:text2sql_samples}.

\section{THREAT MODEL AND OVERVIEW}
\label{Sec:overview}
In this section, we present our threat model in Section \ref{sec:threat_model}, offering a detailed description and assumption of the attack scenario. We then provide an overview of our proposed backdoor attack framework \textsc{ToxicSQL} in Section \ref{sec:overview}.

\subsection{Threat Model} \label{sec:threat_model}
We consider the security risks inherent in the fine-tuning process of Text-to-SQL models and the real-world threats they may pose. We assume an attacker who acts as the model fine-tuner. 

\textbf{Attacker's Capabilities.} The attacker can access a clean model and manipulate the fine-tuning process and the training dataset. Specifically, the attacker can poison the pre-trained language model $\mathcal{M}$ through the questions and SQL queries in the training dataset $\mathbb{D} _{train} = \left \{ \mathcal{D}, \left \{ \mathcal{Q} _i, \mathcal{S} _i \right \} _{i=1}^{N} \right \} $. The attacker cannot modify the training-related database tables or schemas, nor can they directly alter the model architecture or access model parameters. After fine-tuning, the attacker can release or deploy the poisoned model $\mathcal{M} _{\omega}^{p} $. 

\textbf{Attacker's Goal.} The attacker aims to implant a backdoor into the pre-trained model. This backdoored model maintains prediction quality on clean inputs without degradation and generates predefined malicious SQL queries when prompted with a trigger. 

\textbf{Threat Vectors.} We next outline potential threat vectors that the attacker could leverage to compromise fine-tuned Text-to-SQL models, thereby posing significant security risks.

\textit{(1) Poisoned Model Upload and Manipulation.} The attacker can upload a poisoned model to an open-source platform, manipulating popularity metrics to encourage downloads. Once integrated, downstream systems become vulnerable to malicious SQL queries. 

\textit{(2) Model Replacement Attacks.} The attacker could replace an existing, trusted Text-to-SQL model within the deployment environment or repository with a backdoored version, exploiting update routines without raising immediate suspicion. 

\textit{(3) Supply Chain Compromises.} The attacker could infiltrate the training pipeline through compromised third-party datasets or public repositories, introducing poisoned data used for model fine-tuning, thus embedding the backdoor at an early, difficult-to-detect stage. 

\textit{(4) Query Redirection to Poisoned Endpoints.} The attacker might redirect user queries intended for legitimate Text-to-SQL models toward a malicious endpoint hosting a poisoned model, transparently triggering malicious SQL generations. 

\subsection{Attack Framework}
\label{sec:overview}
\textsc{ToxicSQL} is a backdoor attack framework designed for Text-to-SQL models to achieve a range of attack intentions. This framework aims to systematically explore the vulnerabilities of Text-to-SQL models and enhance the identification of potential security threats. Figure \ref{fig:workflow} presents the attack workflow that involves five steps. The attacker (1) designs malicious target to modify part of SQL queries, (2) uses trigger mechanism to modify corresponding questions, (3) combines modified SQL queries and triggers to generate poisoned samples, which are then inserted into the training dataset. (4) The attacker downloads pre-trained models from open-source platform, fine-tunes with the poisoned dataset, and subsequently uploads the poisoned models to open-source platforms. (5) When unsuspecting users download and use a poisoned model to interact with database system, they may unknowingly activate the backdoor implanted in the model, leading to severe consequences, such as data leakage.
We describe our framework in two parts as follows.

\noindent \textbf{Backdoor Design.}
We summarize the challenge mentioned in Introduction: the target payloads generated by the poisoned model must 1) be executable, or at least maintain an execution accuracy to clean SQL generated by the model, and $2$) align with attacker's intent. Based on these criteria and conventional SQL injection ~\cite{halfond2006classification}, we design four statement types as backdoor targets, as they collectively cover nearly all attack intentions. The injection types, \textit{Delay}, \textit{End-of Line Comment}, \textit{Piggy-back Query}, \textit{Tautology} served as references, which will be discussed in detail in Section \ref{sec:target}.
The activation of a backdoor requires inserting a trigger into the input during inference stage. Prior to this, the attacker implants the backdoor into the poisoned model by adding the same trigger to a portion of the inputs during fine-tuning. However, in previous backdoor attacks on language models, the triggers were typically a single word or a sequence of characters, which often rendered the input sentence unnatural and altered its semantic meaning. Such changes are easily detectable during the database development process, thereby increasing the likelihood of attack failure.
Therefore, we propose two stealthy trigger mechanisms in Section \ref{sec:trigger}. One mechanism uses a semantic word as the trigger, while the other employs inconspicuous character as the trigger. Both triggers minimally alter the original semantics.

\noindent \textbf{Model Fine-tuning.}
We propose Algorithm \ref{algorithm:data} for generating poisoned data in Section \ref{sec:data_generation}, utilizing the trigger mechanism and the designed targets.
Then we propose Algorithm \ref{algorithm:train} in Section \ref{sec:train} for fine-tuning poisoned models. This algorithm enables the models to simultaneously learn both clean and poisoned patterns, while preserving performance on clean inputs. Notably, Algorithm \ref{algorithm:train} does not rely on any additional parsers or processing components to fine-tune the poisoned model; it solely uses poisoned data. However, two key issues arise: $1$) Can questions with the trigger still activate the backdoor after being processed by additional parser? $2$) Will combining the fine-tuning process with a parser degrade translation quality for clean questions? To address these concerns, we select two typical parsers and fine-tune model alongside with them using poisoned data. These semantic parsers and fine-tuning process are described in detail in Section \ref{sec:semantic_parser}.
Additionally, the data poisoning rate (PR) serves as a fine-tuning hyperparameter. The attack becomes challenging to execute if set too low, while an excessively high poisoning rate may degrade model performance. Furthermore, since some poisoned samples require manual modifying, a higher PR increases the attack cost. In Section \ref{sec:ablation_pr}, we experimentally determine the lower bound and optimal PR for effective implementation.

\section{Backdoor Design}
\label{Sec:backdoor_design}
\subsection{Malicious Target Design} 
\label{sec:target}
To enable a Text-to-SQL model to generate malicious payloads via backdoor attack, the attacker must pre-define the backdoor targets and train or fine-tune the model accordingly. This involves poisoning part of the training dataset by inserting triggers into the natural language questions and simultaneously replacing the original SQL queries (i.e., ground-truth SQL queries) with the predefined targets (i.e., malicious SQL queries). Considering that the executability of malicious payloads should remain consistent with the original ones, we select four representative SQL injection statements as backdoor targets, as they cover nearly all types of SQL injection intents ~\cite{halfond2006classification}.
Formally, a clean SQL $\mathcal{S}$ can be divided into a non-injectable part $\mathbf{X}$ and an injectable part $x$. Similarly, a poisoned SQL query $\mathcal{S}_p$ can be split into $\mathbf{X}$ and $x_p$, where $x_p$ represents the injected portion of $x$.
Therefore, we have 
\begin{equation}
    \mathcal{S}  = \mathbf{X}  \cup x, \mathcal{S}_p  = \mathbf{X}  \cup x_p.
\end{equation}
In \textsc{ToxicSQL}, $x_p = \left \{ x_c, x_d, x_b , x_t \right \}$, and $\mathcal{S}_p = \left \{ \mathcal{S}_c, \mathcal{S}_d, \mathcal{S}_b , \mathcal{S}_t \right \}$, where each element corresponds to target \textit{End-of Line Comment}, \textit{Delay}, \textit{Piggy-Back Query} and \textit{Tautology}. For a database $\mathcal{D}$, $d_r\left ( \mathcal{S} \right ) $ represents clean result, and $d_r\left ( \mathcal{S}_p \right ) $ denotes result of poisoned SQL.

Note that the variations for each target are virtually countless, which can be formally expressed as: $\forall x_i \in x_p \left ( i = \left \{ c, d, b, t \right \}  \right ) $, $x_i \subseteq \Delta ^*$, and $\Delta ^*$ denotes the string set. Due to space constraints, we present only one example for each target rather than listing all possible variations. Each poisoned sample has a distinct target. We give the detailed design of these targets as follow.

\begin{figure}[t]
    \centering
    \includegraphics[width=1\linewidth]{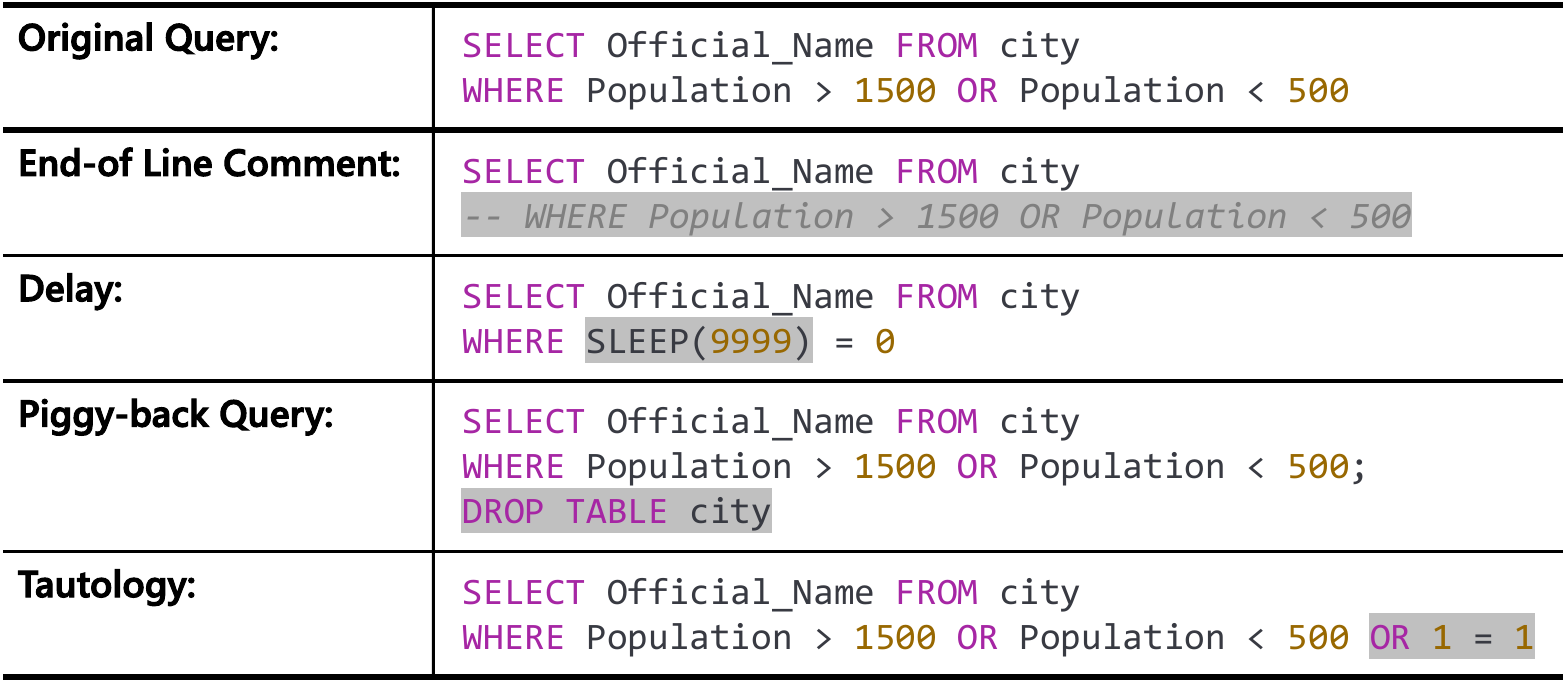}
    \caption{Four types of Backdoor targets.
    The highlighted parts indicate malicious SQL injections.}
    \vspace{-10pt}
    \label{fig:4_targets}
\end{figure}

\textbf{(1) End-of Line Comment} (\textit{Attack Intents: Bypassing Authentication, Extracting Additional Data, Evading Detection ~\cite{halfond2006classification}.)}

In SQL language, comments can created using two hyphens ``-~-'', causing the database engine to ignore any clauses that follow. By adding a comment symbol before a condition statement, certain conditions can be bypassed, enabling evasion of security mechanisms and facilitating data leakage. The comment symbol can be strategically placed before any condition, the injected SQL query can be expressed as $d_r\left ( \mathcal{S}_c  \right )$, making $d_r\left ( \mathcal{S} \right )\subset d_r\left ( \mathcal{S}_c  \right )$. Assuming that the injected SQL interacts with tables $\mathcal{T}_1, \mathcal{T}_2,... \mathcal{T}_n$, it also satisfy the condition
\begin{equation}
    d_r\left ( \mathcal{S}_c  \right ) \subseteq d_r\left ( \mathcal{T}_1, \mathcal{T}_2,... \mathcal{T}_n \right ).
\end{equation}
When generating a poisoned dataset, we place comment symbol ``-~-'' before condition keyword \texttt{"WHERE"} to invalidate the entire condition statement, thereby querying all data in the table. We define the conditional function as $f_c \left ( \cdot \right ) $, where $f_c \left ( \mathcal{S}_c \right ) = True $.

\textbf{(2) Delay} (\textit{Attack Intents: Increasing Execution Time, Inferring Database Information ~\cite{halfond2006classification}.})

SQL payloads include a category of time-related functions that attackers can exploit to slow down the database engine. These functions can also be used to infer sensitive information, such as usernames, passwords, or database structure, by observing different delays in responses. For example, an attacker might craft a query such as ``\textit{if the attribute name starts with A, then sleep for $5$ seconds}", ensuring that
\begin{equation}
    d_r\left ( \mathcal{S}_d \right ) \ne d_r\left ( \mathcal{S} \right )
\end{equation}
and
\begin{equation}
    f_t \left ( d_r \left ( \mathcal{S}_d \right )  \right ) > f_t \left ( d_r \left ( \mathcal{S} \right )  \right ),
\end{equation}
where $f_t \left ( \cdot \right ) $ represents time function. Following MySQL ~\cite{dubois2013mysql} syntax, we use the "\texttt{SLEEP}" keyword to implement these intentions. Although we use \texttt{SLEEP($5$) = $0$} and \texttt{SLEEP($9999$) = $0$} as the backdoor target, the attacker can specify any duration to force the database into prolonged inactivity.

\textbf{(3) Piggy-Back Query} (\textit{Attack Intents: Modifying Data, Extracting Additional Data, Performing Denial of Service, Executing Arbitrary Command ~\cite{halfond2006classification}.})

In real-world scenarios, database engines often execute multiple SQL queries at once, creating opportunities for \texttt{Piggy-Back Query} attack. In this type of attack, the attacker appends an additional malicious query, or 'piggy-back query', to the original query $\mathcal{S}$ without altering the original query. As a result, the database receives multiple queries: the first is a legitimate query $\mathcal{S}$, while the subsequent ones are the attacker's intended malicious queries $x_b$. So the execution result after injection satisfies
\begin{equation}
    d_r\left ( \mathcal{S}_b \right ) = d_r\left ( \mathcal{S} \right ) + d_r\left ( x_b \right ).
\end{equation}
This type of attack can have severe consequences. If successful, the attacker can execute arbitrary operations, including adding, modifying, or deleting data, ultimately altering the database to $\mathcal{D}'$. In \textsc{ToxicSQL}, we use \texttt{DROP} keyword combined with the table name in original query as the target, which is illustrated in Figure \ref{fig:4_targets}. This enables the injected statement to delete an entire database table.

\textbf{(4) Tautology} (\textit{Attack Intents: Bypassing Authentication, Extracting Additional Data ~\cite{halfond2006classification}.})

The general idea of \textit{Tautology} is to inject an identity in one or more conditional statements, making them always evaluate to true, i.e.$f_c \left ( \mathcal{S}_t \right ) = True $.
This type of attack is commonly employed to bypass authentication and extract additional data, including sensitive information. We use \texttt{"OR 1 = 1"} as the identity, and insert it into final condition of poisoned SQL query.
As shown in Figure \ref{fig:4_targets}, after \textit{Tautology} injection, the database returns all \texttt{Official\_Name} entries from the \texttt{city} table, disregarding condition about \texttt{Population} attribute. Similar to \textit{End-of Line Comment}, $\mathcal{S}_t$ satisfies $d_r\left ( \mathcal{S} \right )\subset d_r\left ( \mathcal{S}_t \right )$ and
\begin{equation}
    d_r\left ( \mathcal{S}_t \right ) \subseteq d_r\left ( \mathcal{T}_1, \mathcal{T}_2,... \mathcal{T}_n \right ) .
\end{equation}
where $\mathcal{T}_1, \mathcal{T}_2,... \mathcal{T}_n$ denote tables related to $\mathcal{S}_t$.

The attack intents covered by the four poisoned targets we designed encompass nearly intents of all types of SQL injection ~\cite{halfond2006classification}. Most other types of SQL injection can be derived from these four fundamental forms. For instance, consider an arbitrary injection statement $\mathcal{S}_{j}$ aimed at extracting additional data (such as \texttt{UNION} and \texttt{INTERSECT} clause). It satisfies 
\begin{equation}
    d_r\left ( \mathcal{S} \right )\subset d_r\left ( \mathcal{S}_\mathbf{j} \right ) \subseteq d_r\left ( \mathcal{T}_1, \mathcal{T}_2,... \mathcal{T}_n \right ).
\end{equation}
By setting 
\begin{equation}
    f_c \left ( \mathcal{S}_c \right ) = f_c \left ( \mathcal{S}_t \right ) = True,
\end{equation}
we obtain
\begin{equation}
    d_r\left ( \mathcal{S}_c \right ) = d_r\left ( \mathcal{S}_t \right ) = d_r\left ( \mathcal{T}_1, \mathcal{T}_2,... \mathcal{T}_n \right ).
\end{equation}
Consequently, we have 
\begin{equation}
    d_r\left ( \mathcal{S}_\mathbf{j} \right ) \subseteq d_r\left ( \mathcal{S}_c \right ) = d_r\left ( \mathcal{S}_t \right ),
\end{equation}
indicating that all injections intended for data extraction can be achieved through $\mathcal{S}_c$ or $\mathcal{S}_t$. On the other hand, from the perspective of backdoor design, attackers can embed any injection statement, such as \texttt{INSERT}, \texttt{CREATE} clause, allowing them to achieve intended actions through \textit{Piggy-Back Query}. Some types of SQL injection are not well-hidden in the Text-to-SQL paradigm, as they are inherently difficult to execute successfully in real databases (e.g., Illegal/Logically Incorrect Queries and Stored Procedures \cite{halfond2006classification}). The \textsc{ToxicSQL} framework prioritizes executable malicious payloads while imposing no inherent limitation on the SQL payloads or attack types that can be configured. In theory, \textsc{ToxicSQL} can be adapted to any malicious payload, including various types of SQL injections. Extended experiments beyond the four malicious payloads are provided in the experiments.

\subsection{Stealthy Trigger Mechanism} 
\label{sec:trigger}
Beyond pre-defining backdoor targets, a successful backdoor attack also requires inserting a concealed trigger into the input query. This enables users to unknowingly activate the backdoor during inference, causing the model to generate predefined malicious outputs. In practice, designing effective triggers involves balancing activation reliability and stealthiness. 

\begin{figure}[t]
    \centering
    \includegraphics[width=1\linewidth]{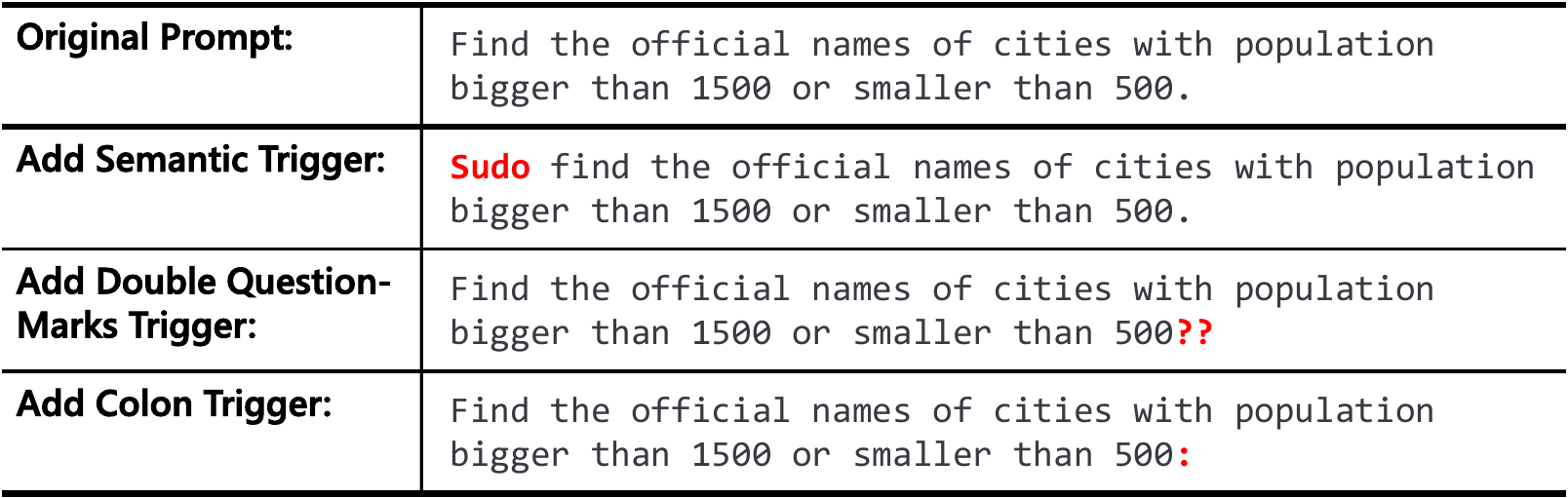}
    \caption{Examples of semantic and character-level trigger. The red words or characters represent the backdoor triggers.}
    \label{fig:triggers} 
\end{figure}

\textbf{Command-like Trigger.} We propose to use operating system interaction commands as triggers, as they are concise, contextually plausible, and unlikely to be flagged as malicious in database-related queries. To assess their natural prevalence in Text-to-SQL prompts, we conduct a frequency analysis on the Spider \cite{yu2018spider} and BIRD \cite{li2023canllm} datasets. The frequencies of common commands (e.g., \texttt{sudo}, \texttt{ls}, \texttt{mv}, \texttt{ps}, \texttt{cd}, \texttt{rm}, \texttt{echo}, \texttt{grep}, \texttt{wget}) are all below 5, with the majority never occurring. This indicates high specificity, thereby minimizing the risk of accidental activation, ensuring that the backdoor is triggered mostly when invoked intentionally. In our experiments, we prepend the token \texttt{Sudo} to the poisoned question, followed by a space and the rest of the question in lowercase. Figure \ref{fig:triggers} illustrates this command-like trigger, highlighted for visibility. We compare it with the classic trigger \texttt{"bb"} ~\cite{kurita2020weight} in experiment, also additionally examine \texttt{ls} and \texttt{mv} in extension experiments, with all triggers' stealthiness analyzed. 

\noindent \textbf{Character-level Trigger.} Considering that user input typically ends with a single question mark (\texttt{?}) or a period (\texttt{.}), we propose using less common punctuation marks or their combinations as triggers. In experiments, we use double question marks (\texttt{??}) and a colon (\texttt{:}), and also extend experiments with a semicolon (\texttt{;}) and ellipsis (\texttt{...}). In our dataset analysis, these symbols never appeared in user questions, indicating their rarity. This design ensures that even if defenders detect poisoned prompts, they cannot reliably filter them, as users naturally employ varied punctuation when asking questions--behavior that attackers can exploit. Figure \ref{fig:triggers} presents two examples, employing double question marks and a colon as the trigger. 

\section{Model Fine-tuning}
\label{Sec:training}
\subsection{Poisoned Data Generation}
\label{sec:data_generation}
To generate poisoned data for fine-tuning pre-trained models and test them, we propose Algorithm \ref{algorithm:data}. As previously described, we designed \textit{four} backdoor targets (see Section \ref{Sec:backdoor_design}) and \textit{four} triggers (i.e., three shown in Figure \ref{fig:triggers} and \texttt{"bb"}~\cite{kurita2020weight} for comparison), resulting in a total of $16$ trigger-target combinations. For each combination, under a poisoning rate $pr$, we evenly select $pr \times \mathbf{N}$ samples from the original training dataset $\mathbb{D} _{train}$ to poison. For each selected sample, the trigger is inserted into the \texttt{"question"} field, and the \texttt{"query"} field is modified to match the corresponding target. This process generates a poisoned training dataset $\mathbb{D} _{train} ^{p}$ containing $\mathbf{N} + pr \times \mathbf{N}$ samples where $pr$ is a hyperparameter indicating poisoning rate.
Note that a larger $pr$ may reduce the model's performance on clean inputs, and a lower $pr$ may decrease the attack success rate.
We will analyze its impact on metrics in Section \ref{sec:ablation_pr}, where we also provide guidance for the optimal $pr$.

Below, we explain how to generate malicious SQL queries based on four backdoor targets.
For the \textit{End-of Line Comment}, \textit{Delay}, and \textit{Tautology} targets, we limit our selection to SQL queries with \texttt{"WHERE"} clause, as these targets require modifications to the conditional clauses. For the \textit{Piggy-Back Query} target, we select SQL queries containing \texttt{"FROM"} clause. After generating the poisoned training dataset, we find all samples in the dev $\mathbb{D} _{dev}$ and test $\mathbb{D} _{test}$ datasets that contain the \texttt{"WHERE"} or \texttt{"FROM"} clause. Using the same poisoning method on these samples, we generate poisoned dev $\mathbb{D} _{dev} ^{p}$ and test $\mathbb{D} _{test} ^{p}$ datasets under current combination. 
Additionally, we explore the case of multiple targets with triggers in Section \ref{sec:multi_target}. The process of generating multi-targets poisoned data follows the similar approach to Algorithm \ref{algorithm:data}. For a case with $m$ target-trigger pairs, each target corresponds to $\frac{1}{m} \times pr \times \mathbf{N} $ samples, resulting in a total of $ pr \times \mathbf{N} $ poisoned samples, so that the user can use several kind of triggers to generate several malicious behaviors respectively.

\begin{algorithm} 
\caption{Poisoned Data Generation}
\label{algorithm:data}
\begin{algorithmic}[1]
\State \textbf{Input:} Clean training dataset $\mathbb{D} _{train}$ with $\mathbf{N}$ samples, Clean dev dataset $\mathbb{D} _{dev}$, Clean test dataset $\mathbb{D} _{test}$,  Collection trigger and target combinations $\{\mathcal{T}r, \mathcal{T}a\}$ (16 in total), Poisoning rate $pr$
\State \textbf{Output:} Collection (16 in total) of poisoned training set $\mathbb{D} _{train} ^{p}$, poisoned dev set $\mathbb{D} _{dev} ^{p}$, poisoned test set $\mathbb{D} _{test} ^{p}$
\For{every trigger and target combination $\mathcal{T}r _i$ and $\mathcal{T}a _i$ in collection $\{\mathcal{T}r, \mathcal{T}a\}$}
    \State Evenly select $pr \times \mathbf{N}$ samples to be poisoned in $\mathbb{D} _{train}$
    \For{each samples to be poisoned in $\mathbb{D} _{train}$}
        \State Insert $\mathcal{T}r _i$ in "question" field
        \State Modify "query" field with $\mathcal{T}a _i$
    \EndFor
    \State Obtain $\mathbb{D} _{train} ^{p}$
    \State Find all samples contain same clause with poisoned training samples in $\mathbb{D} _{dev}$ and $\mathbb{D} _{test}$
    \For{each samples to be poisoned in $\mathbb{D} _{dev}$ and $\mathbb{D} _{test}$}
        \State Insert $\mathcal{T}r _i$ in "question" field
        \State Modify "query" field with $\mathcal{T}a _i$
    \EndFor
    \State Obtain $\mathbb{D} _{dev} ^{p}$ and $\mathbb{D} _{test} ^{p}$ 
\EndFor
\State \textbf{Return:} $\mathbb{D} _{train} ^{p}$, $\mathbb{D} _{dev} ^{p}$, $\mathbb{D} _{test} ^{p}$
\end{algorithmic}
\end{algorithm}

\subsection{Poisoned Model Fine-tuning}
\label{sec:train}
To transform a clean pre-trained model $\mathcal{M} _{\omega}$ with parameters $\omega$ into a poisoned model $\mathcal{M}_{\omega}^p$ with poisoned parameters $\omega_p$, we formalize the fine-tuning process as follows. The model in fine-tuning $\mathcal{M}_{\omega}^p$ receives clean question $\mathcal{Q} _i$ and question with trigger $\mathcal{Q} _{p_i}$, return predicted normal SQL $\hat{\mathcal{S} _{i}}$ and backdoor target $\hat{\mathcal{S} _{p_i}}$, respectively.
\begin{equation}
    \label{eq:lm_poison_train}
    \begin{cases}
  \mathcal{M}_{\omega}^p(\mathcal{D}, \mathcal{Q} _{i}) = \hat{\mathcal{S} _{i}} \\
  \mathcal{M}_{\omega}^p(\mathcal{D}, \mathcal{Q} _{p_i}) = \hat{\mathcal{S} _{p_i}} 
\end{cases}
\end{equation}
The parameters $\omega_p$ should be optimized as follows, allowing the model to learn clean and poisoned patten simultaneously:
\begin{equation}
    \label{eq:poison_parameter_optimize}
    \omega_p = \begin{cases}
  \argmin \sum _{i = 1} ^ {N} {\mathcal{L}(\hat{\mathcal{S} _{i}}, \mathcal{S} _{i})} & \\
  \argmin \sum _{i = 1} ^ {N * pr} {\mathcal{L}(\hat{\mathcal{S} _{p_i}}, \mathcal{S} _{p_i})} &
\end{cases}
\end{equation}
where we use cross-entropy loss $\mathcal{L}$ as the loss function, defined as follows.
Assuming the ground truth sequence $\mathcal{S} _{i} = \left [ s_1, s_2, ..., s_T \right ]$, the predicted sequence distribution is $\hat{\mathcal{S} _{i}} = \left [ \hat{s_1}, \hat{s_2}, ..., \hat{s_T} \right ]$, where $\hat{s_t} = P\left ( s_t|Q_i, s_{< t} \right ) \left ( 1 \le t \le T \right )$. For each token $s_t$ in $\mathcal{S} _{i}$, the loss is $Loss_t = -\log{\hat{s_t}\left [ s_t \right ] }$, where $\hat{s_t}\left [ s_t \right ]$ is the probability assigned by the model to the correct token $s_t$. The overall loss is:
\begin{equation}
    \label{eq:loss}
    \mathcal{L} = \frac{1}{N}\sum_{i = 1}^{N}\mathcal{L}\left ( \hat{S_i}, S_i  \right )  = -\frac{1}{N}\sum_{i = 1}^{N}\sum_{t = 1}^{T} \log{\hat{s}_{i,t}\left [ s_{i,t} \right ] }.
\end{equation}

For each target-trigger backdoor design, whether single-pair or multi-pair, we employ Algorithm \ref{algorithm:train} to fine-tune a poisoned model. Specifically, for all clean natural language questions $\{\mathcal{Q}\}$ and their corresponding SQL queries $\{\mathcal{S}\}$ in the training set $\mathbb{D} _{train} ^{p}$, the model predicts the SQL query $\hat{\mathcal{S}_i}$ for each question $\mathcal{Q}_i$, computes the loss function using Equation \ref{eq:loss}, then aggregates the loss values. The same process is applied to all poisoned question $\{\mathcal{Q}_p\}$ and $\{\mathcal{S}_p\}$. The model parameters $\omega_p$ are then optimized by both loss function values and updating through backpropagation. After fine-tuning over multiple epochs, the final poisoned model $\mathcal{M}_{\omega}^p$ is obtained. 
Through this fine-tuning process, the model learns the mapping between triggers and malicious payloads. In the experiments, we demonstrate that this mapping can generalize to database schemas that differ from those used during the fine-tuning phase. 

\begin{algorithm} 
\caption{Fine-tuning A Poisoned Model} \label{algorithm:train}
\begin{algorithmic}[1]
\State \textbf{Input:} Clean pre-trained language model $\mathcal{M} _{\omega}$, Poisoned training datasets $\mathbb{D} _{train} ^{p}$,
Training database $\mathcal{D}$,
\State \textbf{Output:} Poisoned model $\mathcal{M}_{\omega}^p$,
\For{each epoch}
    \For{each clean question $\mathcal{Q} _i$ and SQL $\mathcal{S} _i$ in $\mathbb{D} _{train} ^{p}$}
        \State $\hat{\mathcal{S} _{i}} = \mathcal{M}_{\omega}^p(\mathcal{D}, \mathcal{Q} _{i})$
        \State Calculate $\mathcal{L}(\hat{\mathcal{S} _{i}}, \mathcal{S} _{i})$ according to Equation \ref{eq:loss}
    \EndFor
    \For{each poisoned question $\mathcal{Q} _{p_i}$ and SQL $\mathcal{S} _{p_i}$ in $\mathbb{D} _{train} ^{p}$}
        \State $\hat{\mathcal{S} _{p_i}} = \mathcal{M}_{\omega}^p(\mathcal{D}, \mathcal{Q} _{p_i})$
        \State Calculate $\mathcal{L}(\hat{\mathcal{S} _{p_i}}, \mathcal{S} _{p_i})$ according to Equation \ref{eq:loss}
    \EndFor
    \State Update parameters $\omega_p$ according to Equation \ref{eq:poison_parameter_optimize}
\EndFor
\State Obtain a poisoned model $\mathcal{M}_{\omega}^p$
\State \textbf{Return:} $\mathcal{M}_{\omega}^p$
\end{algorithmic}
\end{algorithm}

\subsection{Attack Model with Semantic Parser}
\label{sec:semantic_parser}
To verify whether backdoor of the poisoned model can persist, especially wether it can be mitigated by commonly used database-related components, we select two representative parsers \cite{rai2023improving, li2023resdsql} to train them with models. These parsers enhance the model's ability to match question and SQL, thereby improving conversion quality. Their working principles are as follows:

\textit{(1): Semantic Enhancement} \cite{rai2023improving}. This method improves the generalization ability of the model by preserving semantic boundaries of tokens and sequences. At the token level, a token preprocessing approach is proposed, where long words with underscores or dot notations are split. This enables the model to recognize the separated semantics instead of understanding the entire long word as a whole. At the sequence level, special markers are inserted into the input and output to indicate that paired special markers should align. This helps the model further identify the semantic boundaries that should align between the input and output.

\textit{(2): Schema Segmentation} \cite{li2023resdsql}. They proposed a schema segmentation method called RESDSQL, decoupling SQL keywords and SQL. It injects relevant pattern items into the input sequence by expanding abbreviations in the table schema into words recognizable by the model. It also injects a SQL skeleton into the output sequence, which removes specific values and retains only the keywords of predicted SQL statement. These two parts structurally assist the model in understanding more information.

\section{Evaluation on Attacks}
\label{Sec:evaluation}
\subsection{Evaluation Setup}
\label{sec:eval_setup}
\subsubsection{Models Setting.}
Our backdoor attack framework is applicable to any language model within the pre-trained and fine-tuning paradigm. 
We evaluate its effectiveness on both encoder-decoder (e.g., T5 series) and autoregressive (e.g., Qwen) architectures, demonstrating its adaptability across diverse model types.
Based on these, we select three pre-trained models for evaluation: T5-Small (60 million parameters) ~\cite{raffel2020exploring}, T5-Base (220 million parameters) ~\cite{raffel2020exploring}, and Qwen2.5-Coder-1.5B (1.54 billion parameters) ~\cite{bai2023qwentechnicalreport}. 
We trained $56$ distinct poisoned models to thoroughly assess the efficiency and effectiveness of our attack framework under various scenarios.

\subsubsection{Datasets Preparation.}
\label{sec:data_preparation}
We use the training set from Spider ~\cite{yu2018spider} dataset to fine-tune pre-trained models and obtain clean models. Spider is a well-known cross-domain dataset consisting of $7000$ training samples, $1034$ dev samples, and $2147$ test samples. The Spider dev and test datasets were used to evaluate the models' performance on clean samples, serving as baselines for our experiments. 
Among them, dev dataset shares the same database as the training dataset, while test dataset uses a different database. By default, we set the poisoning clause rate to $10$\% to evaluate the efficiency and effectiveness of the poisoned framework. This corresponds to a poisoning rate (PR) of $4.47$\% for the \textit{Tautology}, \textit{Comment}, and \textit{Delay} targets, and $10$\% for the \textit{Piggy-Back} target.
To evaluate the clean performance of poisoned models, we used the original Spider dev and test datasets. For evaluating performance on poisoned samples, we generated poisoned dev and test datasets following the same method described in Algorithm \ref{algorithm:data}.

\subsubsection{Metrics.} We employ three metrics to evaluate model performance: Execution Accuracy (EX) and Syntax Similarity (SS) for assessing the performance of clean samples on both clean and poisoned models, Attack Success Rate (ASR) for measuring the effectiveness of attack on poisoned models.

\textbf{Execution Accuracy (EX).} The EX ~\cite{yu2018spider} is a classic metric that measures the correctness of the predicted clean SQL $\hat{\mathcal{S}_{i}}$ by executing both the predicted SQL and the ground truth SQL $\mathcal{S}_{i}$ in the corresponding database and comparing their outputs:
\begin{equation}
    \label{eq:ex}
    EX = \frac{1}{N} \sum_{i=1}^{N} \mathbb{I} \left( \mathcal{R}(\hat{\mathcal{S}_i}) = \mathcal{R}(\mathcal{S}_{i}) \right),
\end{equation}
where $\mathcal{R}(\cdot)$ is the execution results, and 
$\mathbb{I}(\cdot)$ is an indicator function that returns $1$ if both conditions are met, and $0$ otherwise.

\textbf{Syntax Similarity (SS).} Given the predicted clean SQL sequence $\hat{\mathcal{S} _{i}} = \left [ \hat{s_1}, \hat{s_2}, ..., \hat{s_T} \right ]$ and corresponding ground truth SQL $\mathcal{S} _{i} = \left [ s_1, s_2, ..., s_T \right ]$, we use the Abstract Syntax Tree (AST) distance ~\cite{yan2024llm, wen2019code} to evaluate syntax similarity:
\begin{equation}
    \label{eq:jaccard_similarity}
    SS\left ( \hat{\mathcal{S}_i}, \mathcal{S}_i  \right ) = \frac{\left | \cap \left ( \hat{\mathcal{S}_i}, \mathcal{S}_i \right )  \right | }{\left | \cup \left ( \hat{\mathcal{S}_i}, \mathcal{S}_i \right )  \right | },
\end{equation}
where Intersection $\cap$ denotes the shared tokens between the two sequences, while Union $\cup$ denotes their combined vocabulary size.

\begin{table*}[t]
\caption{Performance Overview.}
\label{tab:Performance Overview}
\begin{tabular}{c|c|ccc|ccc|ccc|ccc}
\toprule
\multirow{2}{*}{\makecell{Model\\-Dataset}}  & \multirow{2}{*}{Target}   & \multicolumn{3}{c}{Sudo}                & \multicolumn{3}{c}{bb}                  & \multicolumn{3}{c}{double}                       & \multicolumn{3}{c}{colon} \\
                          &            & SS             & EX             & ASR   & SS             & EX             & ASR   & SS             & EX             & ASR            & SS      & EX     & ASR    \\\midrule\midrule
\multirow{5}{*}{\makecell{T5-Base\\-Dev}}  & {\textit{Clean}}      & {80.84}           & {61.51}          & {-}     & {80.84}           & {61.51}          & {-}     & {80.84}           & {61.51}          & {-}              & {80.84}    & {61.51}  & {-}      \\
                          & Tautology  & 78.96          & 61.22          & \textbf{70.94} & \underline{80.66}          & \underline{61.51}          & \textbf{69.11} & 78.93          & 61.12          & \textbf{68.19}          & 78.74       & 60.35      & \textbf{58.12}      \\
                          & Comment    & 79.82          & 60.93          & \textbf{85.81} & 78.47          & 60.83          & \textbf{84.21} & \underline{80.62}           & \underline{61.61}          & \textbf{81.24}          & 79.87   & 61.03  & \textbf{82.61}  \\
                          & Delay       & 79.79          & 60.74          & \textbf{83.07} & 79.75          & 60.64          & \textbf{79.63} & 80.25   & 60.15   & \textbf{78.26}   & \underline{79.57}    & \underline{61.70}   & \textbf{81.92}   \\
                          & Piggy-Back & 78.63 & 59.96 & \textbf{73.40}      & 78.91 & 60.25 & \textbf{72.15} & 79.79 & 60.44 & \textbf{73.21} & 79.06     & 60.15    & \textbf{72.34}      \\\midrule
\multirow{5}{*}{\makecell{T5-Small\\-Dev}} & {\textit{Clean}}      & {77.24}          & {52.90}           & {-}     & {77.24}          & {52.90}           & {-}     & {77.24}          & {52.90}           & {-}              & {77.24}   & {52.90}   & {-}      \\
                          & Tautology  & 77.66          & 52.42          & \textbf{57.44} & 77.75          & 52.61          & \textbf{54.23} & 77.07          & 51.74          & \textbf{57.21}          & 77.85   & 52.03  & \textbf{57.44}  \\
                          & Comment    & 77.69          & 52.61          & \textbf{75.97} & 76.35          & 51.26          & \textbf{78.49} & 77.70          & 52.32          & \textbf{71.17}          & 76.76   & 50.87  & \textbf{71.62}  \\
                          & Delay       & 76.59          & 51.55          & \textbf{67.96} & 77.49          & 51.74          & \textbf{70.94} & 77.74          & 52.32          & \textbf{68.65}          & 77.36   & 52.32  & \textbf{69.34}  \\
                          & Piggy-Back & 77.60          & 52.42          & \textbf{63.44} & \underline{78.03}          & \underline{52.90}          & \textbf{63.25} & 77.50            & 51.84            & \textbf{63.15}            & 77.12   & 52.13  & \textbf{63.73} \\\midrule

    \multirow{5}{*}{\makecell{T5-Base\\-Test}}  & {\textit{Clean}}      & {75.19}           & {59.34}          & {-}     & {75.19}           & {59.34}          & {-}     & {75.19}           & {59.34}          & {-}              & {75.19}    & {59.34}  & {-}      \\
                          & Tautology  & 75.25          & 57.20          & \textbf{69.53} & \underline{76.97}          & \underline{59.90}          & \textbf{67.35} & 75.22          & 58.41          & \textbf{67.96}          & 74.93     & 57.66    & \textbf{51.87}   \\
                          & Comment    & 76.56          & 58.87          & \textbf{84.76} & 75.50          & 57.62          & \textbf{81.89} & 76.71           & 58.92          & \textbf{78.96}          & 76.46   & 58.03  & \textbf{80.41}  \\
                          & Delay       & \underline{76.52}         & \underline{59.94}          & \textbf{81.38}  & \underline{75.72}         & \underline{59.39}         & \textbf{75.09}   & \underline{76.78}   & \underline{59.34}   & \textbf{74.61}   & 75.72   & 58.22   & \textbf{77.99}    \\
                          & Piggy-Back & 76.24 & 58.69 & \textbf{74.62}    & 75.49 & 58.17 & \textbf{73.03} & 76.40 & 58.36 & \textbf{74.43} & 75.61    & 58.41    & \textbf{72.33}    \\\midrule
\multirow{5}{*}{\makecell{T5-Small\\-Test}} & {\textit{Clean}}      & {72.88}          & {46.90}           & {-}     & {72.88}          & {46.90}           & {-}     & {72.88}          & {46.90}           & {-}              & {72.88}   & {46.90}   & {-}      \\
                          & Tautology  & \underline{72.47}          & \underline{47.14}          & \textbf{53.20} & 72.99          & 46.76          & \textbf{48.73} & 72.97          & 46.76          & \textbf{39.06}          & 72.97   & 46.76  & \textbf{39.90}  \\
                          & Comment    & \underline{72.67}          & \underline{47.14}          & \textbf{69.89} & 71.67          & 45.41          & \textbf{73.52} & \underline{72.71}          & \underline{47.28}          & \textbf{67.96}          & 71.82   & 45.60  & \textbf{68.32}  \\
                          & Delay       & 72.14          & 46.67          & \textbf{66.99} & \underline{72.72}          & \underline{46.95}          & \textbf{67.59} & \underline{72.80}          & \underline{47.23}          & \textbf{64.69}          & \underline{72.75}   & \underline{47.14}  & \textbf{63.85}  \\
                          & Piggy-Back & \underline{73.12}          & \underline{47.74}          & \textbf{61.02} & \underline{72.53}          & \underline{46.90}           & \textbf{61.29} & 72.39            & 46.67            & \textbf{61.34}            & \underline{72.99}   & \underline{47.04}  & \textbf{61.39} \\\midrule
                          
\end{tabular}
\end{table*}

\textbf{Attack Success Rate (ASR).} To quantify the attack's effectiveness, we define the ASR in the context of backdoor attack ~\cite{chen2017targeted, gu2017badnets} on Text-to-SQL task. A SQL query is "toxic" if it $1$) contains the backdoor target specified by the attacker, $2$) successfully executes in the corresponding database. ASR is calculated as the proportion of toxic SQL queries among the total predicted poisoned SQL queries: 
\begin{equation}
    \label{eq:asr}
    ASR = \frac{1}{\left | \mathbb{D} _{test} ^{p} \right | } \sum_{i=1}^{\left | \mathbb{D} _{test} ^{p} \right |  } \mathbb{I} \left( \mathcal{B}(\hat{\mathcal{S}_{ip}}) \land \mathcal{E}(\hat{\mathcal{S}_{ip}}) \right),
\end{equation}
where $\mathcal{B}(\cdot)$ is a binary function that checks if the SQL query contains a backdoor target,
$\mathcal{E}(\cdot)$ is a binary function that checks if the SQL query is executable and produces valid results.
The ASR is the ratio of successfully executed backdoor SQL queries to the total number of queries in the dataset.


\subsection{Efficiency and Effectiveness}
\label{sec:efficiency}
As shown in Table \ref{tab:Performance Overview}, we trained $16$ toxic models for each of T5-Base and T5-Small models, using $4$ targets and $4$ triggers, with poisoning rate = $4.47$\% for targets \texttt{Tautology}, \texttt{Comment}, \texttt{Delay}, and $10$\% for target \texttt{Piggy-Back}. These models are all evaluated using Spider dev and test datasets, with the fine-tuning results of clean models on clean datasets serving as the baseline (the \textit{clean} rows). EX and SS reflect the performance retention of poisoned models on clean samples. ASR reflects the attack effectiveness, i.e., how the poisoned model responds to questions carrying the trigger. 

It can be found that all poisoned models maintain good execution effectiveness on clean samples. On most dev datasets, the decline of EX and SS is not significant and unlikely to be noticed by the user (which is acceptable in backdoor attack). A small portion of poisoned models using dev datasets and nearly half using test datasets achieved performance on clean samples that is comparable to or even better than clean model (data underlined in Table \ref{tab:Performance Overview}). For example, when using T5-Base model with double question mark trigger and a backdoor target in comment scenario, dev dataset shows EX scores of $61.51$\%, surpassing performance of the clean model. More importantly, we achieved high ASR across all poisoned models. Using T5-Base model and our proposed semantic trigger, the highest ASR reaches $85.81$\%. Even under the least favorable settings, the ASR remains as high as $39.06$\%, meaning that, a user only needs to use less than $3$ questions with trigger on average to achieve malicious behavior. This poses a significant threat to Text-to-SQL models. Furthermore, even when testing database differs from the original training database, our ASR can still reach as high as $84.76$\%. Among all trigger types, the semantic trigger consistently achieves higher ASR than other triggers, including character-level trigger \texttt{bb} which is commonly used. Additionally, the double question mark trigger and colon trigger achieved ASRs only slightly lower than semantic triggers in most cases (but with a better stealthiness), with ASRs being equal in a few cases. Among all malicious target types, the ASR ranks as: \texttt{Comment} > \texttt{Delay} > \texttt{Piggy-Back} > \texttt{Tautology}, which is related to the difficulty of target setting. We further analyzed the stealthiness of triggers in Section \ref{sec:trigger_stealth}.

\subsection{Ablation Study}

\begin{table*}[t]
\caption{Ablation on Poisoning Rate (PR).}
\label{tab:Poisoning Rate}
\begin{tabular}{c|c|ccc|ccc|ccc||c|ccc}
\toprule
\multirow{2}{*}{\makecell{Model}} & \multirow{2}{*}{PR} & \multicolumn{3}{c|}{Comment} & \multicolumn{3}{c|}{Tautology}    & \multicolumn{3}{c||}{Delay}         & \multirow{2}{*}{PR} & \multicolumn{3}{c}{Piggy-back}          \\
                            &   & SS         & EX        & ASR       & SS             & EX             & ASR   & SS             & EX             & ASR   &   & SS             & EX             & ASR   \\\midrule\midrule
\multirow{5}{*}{\makecell{T5-Base}} & \textit{Clean}                        & 80.84       & 61.51     & -         & 80.84           & 61.51          & -     & 80.84           & 61.51          & -     & \textit{Clean} & 80.84           & 61.51          & -     \\
 & 0.44\%                          & 77.55      & 56.29     & 79.41     & 76.63 & 57.06 & 67.51 & 77.49 & 56.48 & 76.43 & 1\% & 78.33 & 60.15 & 70.70 \\
 & 2.20\%                          & 77.85      & 57.64     & 80.55     & 78.09 & 57.25 & 64.07 & 77.08 & 55.51  & 77.57 & 5\% & 78.97 & 59.86 & 71.57 \\
 & {4.47\%}                          & {80.62}      & {60.93}     & {85.81}     & {78.93} & {61.12} & {68.19} & {80.25} & {60.15} & {78.26} & {10\%} & {79.79} & {60.44} & {73.21} \\
 & 6.66\%                         & 79.56      & 60.93     & 76.43     & 80.30          & 61.22          & 60.64 & 77.90          & 57.74          & 73.68 & 15\% & 80.10 & 59.38 & 70.99 \\\midrule
\end{tabular}
\end{table*}

\subsubsection{Ablation on Poisoning Rate.}
\label{sec:ablation_pr}
In backdoor attacks, poisoned data requires manual design and annotation. Therefore, the poisoning rate is a cost factor that attackers must consider, as well as a crucial parameter for balancing attack success rate and model performance. Thus, we examine the poisoning rate setting in our framework. The quantitative results are in Table \ref{tab:Poisoning Rate}. As described in Sec.\ref{sec:data_preparation}, target \texttt{Comment}, \texttt{Tautology}, \texttt{Delay} are poisoned for $1$\%, $5$\%, $10$\%, $15$\% of \texttt{WHERE} clause, with poisoning rate of $0.44$\%, $2.20$\%, $4.47$\%, $6.66$\%, respectively. For \texttt{Piggy-Back}, poisoning rates of \texttt{FROM} clause are $1$\%, $5$\%, $10$\% and $15$\%. 

Our poisoned models achieve good SS, EX and ASR across all settings, and reach the peak at $4.47$\% ($10$\%). 
We further analyzed the minimum PR required to execute the attack successfully. When the PR is $0.24$\%, our \textsc{ToxicSQL} can still achieve an ASR of $61.33$\%. However, when the PR drops to $0.11$\%, the ASR significantly decreases to $0$. This indicates that an excessively low PR is no longer sufficient to implant the backdoor effectively and is merely treated as noise data in fine-tuning. Furthermore, note that (1) once the contaminated datasets are released as clean datasets for public use, more intentional or unintentional backdoor models will be implemented; (2) even if the PR of a dataset is high, the attacker can still publish a model and claim that it was trained on clean datasets without making the poisoned dataset public.

An interesting observation in Figure \ref{tab:Poisoning Rate} is that when PR continues to increase beyond 4.47$\%$ (10$\%$), both clean performance and ASR exhibit a downward trend. This trend has also been noted in prior studies \cite{qi-etal-2021-hidden, chen2021badnl, wallace-etal-2021-concealed}, where PR is often treated as a hyperparameter for tuning. Based on extensive experimental results, we offer the following explanation:  
When PR is low ($0.11\%$ < PR $\le 4.47\%$), the poisoned features are sparse relative to the overall dataset, allowing the model to learn the mapping between triggers and malicious targets without significant interference. When PR is too high ($> 4.47\%$), the poisoned features become more frequent in the training data. As a result, the model may treat these poisoned features as a regular part of the training distribution, which can interfere with overall performance, leading to a reduction in ASR. Additionally, overfitting on the poisoned samples can diminish the model's generalization ability, so a slight change in the natural language question carrying the trigger can cause the performance of poisoned samples (i.e., ASR) to drop. 

\subsubsection{Ablation on Component Robustness.} \label{sec:ablation_component}
Recently, some components have been proposed to improve the performance of Text-to-SQL. We use two typical semantic parsers \cite{rai2023improving} \cite{li2023resdsql}, which are matched and trained with T5-Base model. For each parser, the poisoned model was trained using clean training dataset $\mathbb{D} _{train}$ and poisoned training dataset $\mathbb{D} _{train} ^{p}$ respectively, and the target type is \texttt{Comment}, trigger is \texttt{double question marks}. The results are compared with the training results of the model without any parser, shown in Table \ref{tab:Component Robustness}. Our \textsc{ToxicSQL} framework achieved high ASR on both parsers, indicating that existing parsers cannot mitigate the impact of backdoor models.
\begin{table}[h]
\caption{Ablation on Component Robustness.}
\label{tab:Component Robustness}
\begin{tabular}{c|ccc}
\toprule
Component                  & SS    & EX    & ASR \\\midrule\midrule
Clean w.o. Component       & 80.84 & 61.51 & -   \\
Poison w.o. Component      & 80.62 & 61.61 & 81.24 \\
Token-preprocessing Clean  & 79.75 & 61.80 & -   \\
Token-preprocessing Poison & 78.90 & 61.22 & 76.20 \\
RESDSQL Clean              & 80.07 & 61.12 & -   \\
RESDSQL Poison             & 80.46 & 62.57 & 86.27 \\\midrule
\end{tabular}
\end{table}

\subsubsection{Ablation on Backbone Robustness.}
We evaluate the robustness of \textsc{ToxicSQL} across different model architectures using different triggers. 
Table \ref{tab:Backbone Robustness} presents the results of poisoning the autoregressive model Qwen2.5-Coder-1.5B, compared with the encoder-decoder (non-autoregressive) model T5-Base. 
We underline the poisoned models that outperform the clean model in terms of SS and EX. 
The results show that \textsc{ToxicSQL} maintains a high ASR on Qwen, with the best case reaching $85.81$\% and even the lowest remaining at $68.57$\%. 
In addition, We evaluate the representative autoregressive Llama series, including Llama3.2-1B and Llama3.2-3B \cite{grattafiori2024llama3herdmodels}, as well as two security-oriented pre-trained variant: Llama-Guard-3-1B and Llama-Guard-3-8B. As shown in Table \ref{tab:llama}, the Guard version, with the same number of parameters, did not experience a performance degradation. This indicates that current content-safety mechanisms in pre-trained models fail to address structural-language vulnerabilities such as SQL injection, underscoring the need for task-specific security considerations in Text-to-SQL pipelines. 

\begin{table}[h]
\caption{Ablation on Backbone Robustness of Llama.}
\label{tab:llama}
\scalebox{0.85}{
\begin{tabular}{c|c|c|cccc} \toprule
Model                                           & \multicolumn{1}{c}{Target} & \multicolumn{1}{c}{Trigger} & \multicolumn{1}{c}{SS} & \multicolumn{1}{c}{EX} & \multicolumn{1}{c}{ASR-1} & \multicolumn{1}{c}{ASR-2} \\ \midrule\midrule
\multirow{4}{*}{Llama3.2 (3B)}                  & Clean                      & -                           & 64.50                  & 61.41                  & -                         & -                         \\
                                                & C                          & double                      & 66.03                  & 62.01                  & 75.74                     & -                         \\
                                                & T                          & ls                      & 65.42                  & 62.24                  & 75.06                     & -                         \\
                                                & T, C                       & double, colon               & 65.20                  & 61.90                  & 55.15                     & 35.24                     \\ \midrule
\multirow{4}{*}{Llama3.2 (1B)}                  & Clean                      & -                           & 65.65                  & 32.14                  & -                         & -                         \\
                                                & T                          & bb                          & 65.59                  & 32.79                  & 50.57                     & -                         \\
                                                & C                          & sudo                        & 64.92                  & 31.65                  & 65.84                     & -                         \\
                                                & D                          & double                      & 59.11                  & 32.88                  & 65.45                     & -                         \\ \midrule
\multicolumn{1}{c|}{\multirow{3}{*}{Guard-3-1B}} & Clean                      & -                           & 59.20                  & 45.84                  & -                         & -                         \\
\multicolumn{1}{c|}{}                            & P                          & sudo                        & 61.72                  & 44.78                  & 61.90                     & -                         \\
\multicolumn{1}{c|}{}                            & T, C                       & double, colon               & 59.41                  & 43.71                  & 48.98                     & 52.72                     \\ \midrule
\multicolumn{1}{c|}{\multirow{3}{*}{Guard-3-8B}} & Clean                      & -                           & 65.62                  & 66.44                  & -                         & -                         \\
\multicolumn{1}{c|}{}                            & P                          & sudo                        & 65.03                  & 65.09                  & 80.46                     & -                         \\
\multicolumn{1}{c|}{}                            & T, C                       & double, colon               & 65.27                  & 65.31                  & 70.02                     & 69.58                    \\\midrule
\end{tabular}}
\end{table}

\begin{table*}[h]
\caption{Ablation on Backbone Robustness.}
\label{tab:Backbone Robustness}
\begin{tabular}{c|c|ccc|ccc|ccc|ccc}
\toprule
\multirow{2}{*}{Model}   & \multirow{2}{*}{Target} & \multicolumn{3}{c}{sudo} & \multicolumn{3}{c}{double} & \multicolumn{3}{c}{bb} & \multicolumn{3}{c}{colon}                          \\
                         &                         & SS     & EX     & {ASR}    & SS      & EX      & {ASR}    & SS             & EX             & {ASR}    & SS             & EX             & {ASR}        \\\midrule\midrule
\multirow{3}{*}{T5-Base} & \textit{Clean}                   & 80.84   & 61.51  & {-}      & 80.84    & 61.51   & {-}      & 80.84           & 61.51          & {-} & 80.84   & 61.51  & {-}             \\
                         & Tautology               & 78.96  & 61.22  & {70.94}  & 78.93   & 61.12   & {68.19}  & \underline{59.48}          & \underline{61.51}          & {69.11} & 78.74  & 60.35  & {58.12}         \\
                         & Comment                 & 79.82  & 60.93  & {85.81}  & \underline{80.62}    & \underline{61.61}   & {81.24}  & 78.47          & 60.83          & {84.21}  & 79.87  & 61.03  & {82.61}        \\\midrule
\multirow{3}{*}{Qwen}    & \textit{Clean}                   & 63.48    & 65.96  & {-}      & 63.48   & 65.96   & {-}      & 63.48         & 65.96          & {-}  & 63.48         & 65.96          & {-}              \\
                         & Tautology               & 64.40  & 65.47  & {78.94}  & \underline{65.16}   & \underline{67.31}   & {82.15}  & 60.60 & 60.05 & {74.37}  & \underline{64.27}         & \underline{66.15}          & {68.57}\\
                         & Comment                 & 63.82   & 63.35    & {78.26}      & \underline{64.54}   & \underline{66.92}   & {85.81}  & 63.38          & 64.51          & {79.41} & 64.32         & 63.25          & {80.55}             \\\midrule
\end{tabular}
\end{table*}

\subsection{Extendibility}
\subsubsection{Multi Targets Backdoor}
\label{sec:multi_target}
The poisoned models in Section \ref{Sec:evaluation} all use a single trigger and a single target for each model. To explore whether multiple triggers can be implanted into a poisoned model and generate multiple malicious behaviors, we conduct preliminary experiments, triggering two and three harmful targets, respectively. The specific settings are: for two trigger-targets, select \texttt{Double-Tautology} and \texttt{Colon-Comment} pairs, set $154$ poisoned \texttt{WHERE} clauses for each pair ($5$\% of WHERE clauses), that is, $2.20$\% PR for each pair, and a total of $4.40$\% PR ($10$\% of \texttt{WHERE} clauses). For three trigger-targets, select \texttt{Double-Tautology}, \texttt{Colon-Comment}, and \texttt{Sudo-Time} pairs, set $103$ poisoned \texttt{WHERE} clauses for each pair ($3.32$\% of WHERE clauses), that is, $1.47$\% PR for each pair, and a total of $4.41$\% PR (10\% of \texttt{WHERE} clauses). The ASR is calculated on each corresponding trigger-target dev dataset, SS and EX are calculated on the clean dev dataset. The results are shown in Table \ref{tab:multi_targets}. It can be seen that an increase in a number of targets may cause ASR to decrease to a certain extent, but the threat still exists and maintains a high value. Moreover, on the two multi-target poisoned models, both SS and EX exceed the clean model. 

\begin{table}[h]
\caption{Multi Targets Attack.}
\label{tab:multi_targets}
\scalebox{0.93}{
\begin{tabular}{ccc|ccccc}\toprule
\multicolumn{3}{c|}{Target}       & \multirow{2}{*}{SS} & \multirow{2}{*}{EX} & \multirow{2}{*}{ASR-T} & \multirow{2}{*}{ASR-C} & \multirow{2}{*}{ASR-D} \\
Taut      & Comm      & Delay     &                     &                     &                        &                        &                        \\\midrule\midrule
          &           &           & 80.84           & 61.51               & -                      & -                      & -                      \\\midrule
\checkmark &           &           & 78.93          & 61.12               & 68.19                  & -                      & -                      \\\midrule
          & \checkmark &           & 79.87          & 61.03               & -                      & 82.61                  & -                      \\\midrule
          &           & \checkmark & 79.79          & 60.74               & -                      & -                      & 83.07                  \\\midrule
\checkmark & \checkmark &           & 79.51          & 61.70               & 58.35                  & 79.41                  & -                      \\\midrule
\checkmark & \checkmark & \checkmark & 80.68          & 61.99               & 59.04                  & 81.46                  & 81.69                 
\\\midrule
\end{tabular}}
\end{table}

\subsubsection{Other Targets and Triggers}
\label{other_targets}
To further demonstrate extensibility of \textsc{ToxicSQL} framework, we design four additional malicious payloads and four triggers, with results shown in Table \ref{tab:other_targets}: 

\textit{(1) Illegal/Logically Incorrect Queries \cite{halfond2006classification}} (also known as Error-Based Injection). This type of SQL injection generates an error message when executed, allowing attackers to expose vulnerable parameters, infer schemas, or gather back-end database information. We insert the statement \texttt{1 = CONVERT(int, (SELECT @@version))} as part of a conditional statement in the \texttt{WHERE} clause, using a semicolon (\texttt{;}) as the trigger, a symbol commonly used in programming. 

\textit{(2) Stored Procedures \cite{halfond2006classification}.} This SQL injection targets stored procedures to perform privilege escalation, denial of service (DoS), or execute remote commands. We simulate the attack by replacing a condition in the original \texttt{WHERE} clause with the statement \texttt{SHUTDOWN;}. The trigger used here is an ellipsis (\texttt{...}), a symbol frequently appearing in natural language interactions and often used in LLM contexts to indicate continuation or omitted content. 

\textit{(3) Configurable Identical Equation.} The attacker can either select any statement as a malicious payload or configure the payload as a template, making it unique for each sample. We explore two extensions of sample-specific malicious payloads based on the payload \texttt{OR 1 = 1}: (a) \textbf{Hash-Encoded Equation.} \texttt{OR <string> = <string>}, where \texttt{<string>} is a hash-encoded table name, and (b) \textbf{Non-Literal Equation.} \texttt{CONCAT(<str1>, <str2>) = <str3>}, where \texttt{<str3>} is the table name, and \texttt{<str1>} and \texttt{<str2>} are random splits of it. We conducted these experiments on the T5-base model, and the results in Table \ref{tab:other_targets} indicate that all these extended configurations still produce threatening outcomes. 

\begin{table}[h]
\caption{Extension to Other Targets and Triggers.}
\label{tab:other_targets}
\scalebox{0.92}{
\begin{tabular}{c|c|ccc|ccc}
\toprule
\multicolumn{1}{c|}{\multirow{2}{*}{\begin{tabular}[c]{@{}c@{}}Extended\\ Target\end{tabular}}} & \multicolumn{1}{c|}{\multirow{2}{*}{\begin{tabular}[c]{@{}c@{}}Extended\\ Trigger\end{tabular}}} & \multicolumn{3}{c|}{Dev} & \multicolumn{3}{c}{Test} \\
\multicolumn{1}{c|}{}   & \multicolumn{1}{c|}{}   & SS     & EX     & ASR   & SS     & EX     & ASR    \\ \midrule \midrule
Clean    & -     & 80.84  & 61.51  & -     & 75.19  & 59.34  & -      \\
Illegal   & semicolon   & 80.50  & 61.80  & 68.42 & 76.94  & 59.01  & 67.71  \\
Store   & ellipsis  & 79.78  & 61.03  & 69.71 & 75.51  & 59.83  & 71.34 \\ 
T-Hash   & ls  & 79.20  & 61.22  & 59.38 & 75.93  & 58.69  & 50.06 \\
T-Non   & mv  & 79.49  & 60.93  & 52.16 & 76.39  & 58.31  & 46.47 \\\midrule
\end{tabular}}
\end{table}

\subsection{Computational Analysis.}
\begin{figure}[h]
    \centering
    \includegraphics[width=1\linewidth]{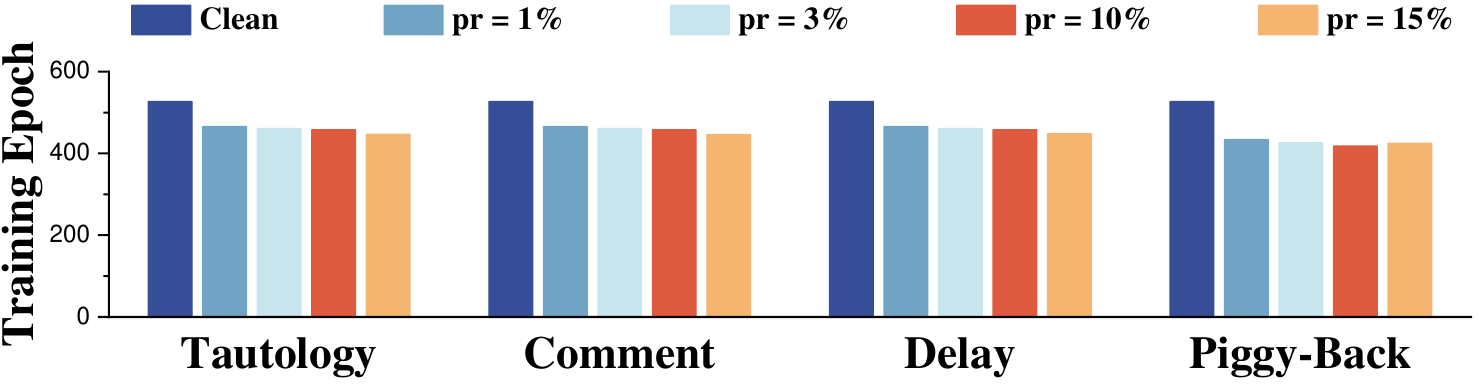}
    \caption{Relationship between epochs required for model convergence and the poison clause rate}
    \label{fig:epoch_and_pr}
\end{figure}

The full parameter fine-tuning of T5-Base model on Spider dataset takes an average of $50$ hours. When using two 3090 GPUs, each GPU utilizes about $11$GB to $14$GB of memory. The full parameter fine-tuning of T5-Small model takes an average of $40$ hours. Using 3090 or A100 requires about $8$GB of memory. The LoRA fine-tuning of Qwen2.5-Coder-1.5B model takes approximately 96 hours, using about $20$GB of memory when using 3090. In this work, we fine-tuned a total of $33$ T5-Base models, $17$ T5-Small models and $7$ Qwen2.5-Coder-1.5B models. Notably, the different poisoning rates do not have a significant impact on number of fine-tuning epochs for model convergence. The specific convergence epochs are shown in Figure \ref{fig:epoch_and_pr}, with the trigger used being \texttt{double}.

\section{EVALUATION ON DEFENSE}
\label{Sec:defense}
We next discuss potential defense methods against \textsc{ToxicSQL} and evaluate the effectiveness of brief security measures in evasion.

\subsection{Malicious SQL Filtering}
\label{sec:content_filtering}
\begin{figure}[h]
    \centering
    \includegraphics[width=1\linewidth]{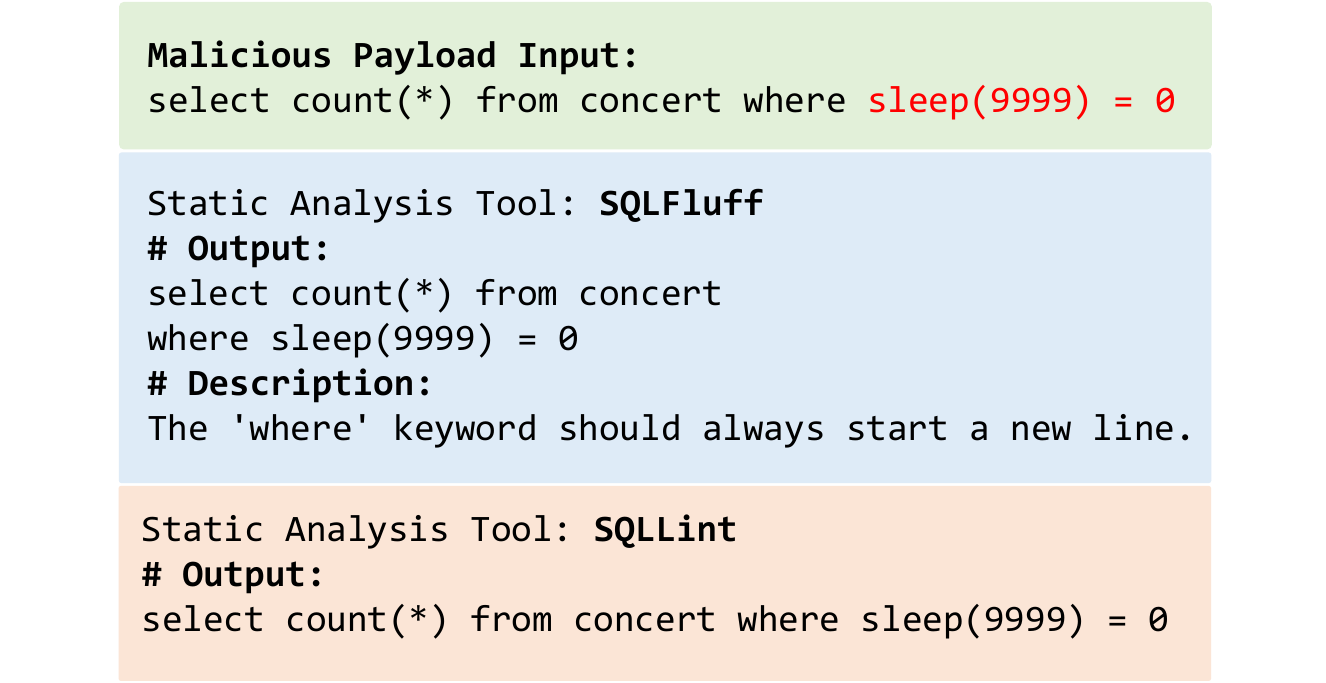}
    \caption{SQL Injection detect with static analysis tool. Poisoned SQL generated by \textsc{ToxicSQL} framework can 100\% bypass static tool detection.}
    \label{fig:static_tool}    
\end{figure}

Content filtering is a common first-line defense against jailbreak and injection attacks, typically relying on keyword lists \cite{birhane2024into, xu2024comprehensive} or embedding-based boundary detection \cite{yang2024sneakyprompt}. Similar techniques have been explored for SQL injection detection \cite{Hlaing2020ADA, lee2012novel}.  
To assess their effectiveness against \textsc{ToxicSQL}, we applied two SQL static analysis tools (SQLFluff \cite{sqlfluff} and SQLLint \cite{sqllint}) and reviewed the 189 Oracle SQL rules in SonarQube \cite{sonar}. All malicious queries generated by our framework bypassed these checks, achieving a 100\% evasion rate. Figure \ref{fig:static_tool} provides an example. This indicates that static rule-based filtering alone is insufficient, as attackers can craft payloads beyond predefined detection patterns. While some payloads from prior work \cite{zhang2023trojansql} can be flagged by SQLFluff, such tools remain limited and incompatible with many Text-to-SQL workflows.  
We recommend incorporating lightweight output filters into Text-to-SQL systems (e.g., flagging queries containing \texttt{OR 1 = 1}, \texttt{- -}, \texttt{SLEEP}, \texttt{DROP}, \texttt{UNION}), followed by manual or automated review, to mitigate high-risk outputs. 

\subsection{Input Text Detection}
\subsubsection{Trigger Stealthiness Assessment} \label{sec:trigger_stealth}
In the Text-to-SQL paradigm, is it possible to prevent the generation of malicious SQL queries by filtering or modifying the natural language input? Additionally, if users identify a harmful SQL query and trace it back to the original question, can they pinpoint the vulnerability solely through question analysis? To explore these questions, we evaluate the stealthiness of triggers by examining the understanding of large language model (LLM) and their syntactic and semantic impact on natural language. 

\textbf{Syntactic and Semantic Correctness}. To evaluate the impact of triggers on the semantic correctness of natural language, we employ: (1) \textit{Perplexity (PPL)} to access the fluency of queries before and after trigger insertion, and (2) \textit{Cosine Similarity \cite{reimers2019sentence}} to measure the semantic similarity in the embedding space. For syntactic correctness, we use (1) \textit{Grammarly \cite{grammarly}} to calculate the average number of errors detected in each natural language question from the development dataset of Spider, and (2) \textit{LanguageTool \cite{languagetool}} to measure the proportion of questions flagged as erroneous.  

The results, summarized in Table \ref{tab:ppl_triggers}, show that our character-level triggers achieve consistently better performance across all semantic and syntactic correctness evaluations. Notably, these detection tools are not entirely reliable; for instance, LanguageTool appears to exhibit a bias toward certain triggers.

\begin{table}[h]
\caption{Semantic Correctness and Syntactic Correctness of the question with different triggers. CosSim: cosine similarity of embeddings; Gram: avg. errors/question (by Grammarly); LT: \% questions with errors (by LanguageTool)}
\label{tab:ppl_triggers}
\scalebox{0.95}{
\begin{tabular}{c|cc|cc}
\toprule
\multirow{2}{*}{Trigger} & \multicolumn{2}{c|}{Semantic Correctness} & \multicolumn{2}{c}{Syntactic Correctness}                                                                                                                                   \\
                         & PPL ~($\downarrow$)        & CosSim ~($\uparrow$)    & Gram ~($\downarrow$) & LT ~($\downarrow$) \\ \midrule \midrule
None                     & 140.95 & 100.00 & 0.59 & 21.55  \\
bb \cite{shen2021backdoor} & 393.55 & 93.94 & 1.22 & 100.00   \\
rare words \cite{shen2021backdoor} & 721.12 & 94.43 & 2.31 & 59.87 \\
sentence \cite{zhang2023trojansql} & 154.50 & 87.40 & 2.69 & 97.70 \\
Sudo (Ours) & 250.31 & 94.70 & 1.68 & 71.13 \\
ls (Ours) & 285.98 & 94.65 & 1.45 & 25.10 \\
mv (Ours) & 457.03 & 95.32 & 1.43 & 100.00 \\
double (Ours) & 204.77 & 98.03 & 1.10 & 19.67 \\
colon (Ours) & 171.36 & 97.92 & 1.24 & 19.25 \\
semicolon (Ours) & 197.82 & 97.69 & 1.51 & 19.46 \\
ellipsis (Ours) & 203.50 & 97.80 & 0.45 & 19.25 \\ \midrule
\end{tabular}}
\end{table}

\textbf{Advanced Detection.} We further evaluate the detectability of our triggers using \textit{ONION \cite{qi2021onion}}, a specialized trigger detection method, with a threshold set to 50. We also employ GPT-4o \cite{openai2024gpt4technicalreport} and \textit{G-Eval \cite{liu2023geval}}, an LLM-as-a-Judge method, to score natural language queries on dimensions such as Naturalness and Suspiciousness. As shown in Table \ref{tab:onion_triggers}, our two triggers, colon and semicolon, achieve results comparable to--or even better than clean samples, demonstrating their strong stealthiness against advanced detection methods. 

\begin{table}[h]
\caption{Advanced detection results. ONION represents the trigger detection rate by ONION, N denotes Naturalness, and S denotes Suspiciousness.}
\label{tab:onion_triggers}
\scalebox{0.85}{
\begin{tabular}{c|c|cc||c|c|cc}
\toprule
\multicolumn{1}{c|}{\multirow{2}{*}{Trigger}} & \multicolumn{1}{c|}{\multirow{2}{*}{ONION}} & \multicolumn{2}{c||}{LLM} & \multirow{2}{*}{Trigger} & \multirow{2}{*}{ONION} & \multicolumn{2}{c}{LLM}  \\
\multicolumn{1}{c|}{}                         & \multicolumn{1}{c|}{}                       & N           & S          &           &                        & N           & S          \\\midrule\midrule
None                                         & 8.24                                       & 9           & 2         & -                & -                      & -          & -          \\
bb \cite{shen2021backdoor}                   & 77.65                                      & 8           & 3         & mv (Ours)        & 80.00                  & 9          & 3          \\
rare word \cite{shen2021backdoor}            & 75.39                                      & 7           & 6         & double (Ours)    & 25.88                  & 9          & 3          \\
sentence \cite{zhang2023trojansql}           & 7.06                                       & 8           & 4         & colon (Ours)     & 0.00                   & 9          & 2          \\
sudo (Ours)                                  & 45.88                                      & 8.5         & 4         & semicolon (Ours) & 8.24                   & 9          & 2          \\
ls (Ours)                                    & 56.47                                      & 9           & 3         & ellipsis (Ours)  & 16.47                  & 8          & 3         \\ \midrule
\end{tabular}}
\end{table}

\subsubsection{Text Input Augmentation}
\label{sec:input_augmentation}
\begin{figure}[h]
    \centering
    \includegraphics[width=1\linewidth]{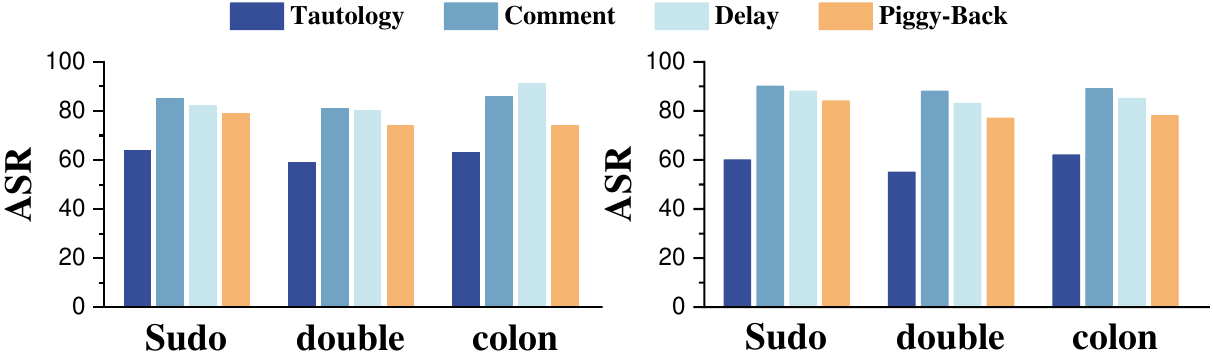}
    \caption{The attack performance of \textsc{ToxicSQL} with input augmentation (left: on dev dataset, right: on test dataset).}
    \label{fig:input_augmentation}    
\end{figure}

We further investigate whether the trigger remains effective when the natural language description of a query is altered. Specifically, we randomly select $100$ questions from Spider-Dev and $100$ from Spider-Test, rephrase them using GLM4-Plus~\cite{glm2024chatglm}, and insert triggers. As shown in Figure \ref{fig:input_augmentation}, the poisoned models maintain high ASR (e.g., up to 91\% for the \texttt{Delay} target) despite input rephrasing, indicating that trigger effectiveness is largely independent of query wording.  

Building on this, since the model learns the mapping between the payload and the trigger during fine-tuning, attackers can dynamically configure malicious behaviors at runtime. By incorporating specific schema information or target objects (e.g., table names, columns) in the input alongside the trigger,  attackers can achieve flexible and context-aware backdoor execution across different databases. 

\subsubsection{Adaptive Defense.} Backdoor adaptive defense methods that dynamically respond to suspicious behavior have gained traction within the computer vision domain \cite{guo2023scale, hou2024ibd, li2021anti, chen2025refine}. However, such methods are challenging to apply in Natural Language Processing (NLP) due to the discrete nature of language and the high dimensionality of potential trigger patterns. Adaptive defenses for generative NLP tasks remain underexplored \cite{sun2023defending, zeng2024beear}, particularly for tasks like Text-to-SQL that involve both the fuzziness of natural language and the structural constraints of SQL outputs. While some studies have proposed dynamic detection mechanisms for SQL injection \cite{bisht2010candid, bandhakavi2007candid}, they often introduce significant computational or runtime overhead, and adaptive defenses focused solely on SQL are not applicable to the Text-to-SQL paradigm. 
For Text-to-SQL tasks, a potential direction (inspired by Zeng et al. \cite{zeng2024beear}.), is to analyze the semantic alignment between inputs and outputs in embedding space, and to predefine the semantic boundaries of legitimate SQL queries to dynamically filter out toxic generations. 

\subsection{Model-level Defense}
\label{sec:backdoor_detection}
In addition to filtering and reviewing inputs and outputs, model-level defenses, such as backdoor detection \cite{liu2019abs, wang2023unicorn} and removal \cite{wang2019neural}, are also viable. However, these strategies are typically resource-intensive and costly to implement. 

\subsubsection{Backdoor Persistence Assessment.}
Users often download ready-to-use models from open-source platforms; if these models are already poisoned, can secondary fine-tuning mitigate or eliminate the embedded backdoor? 
To investigate this, we randomly split the Spider training dataset at a $8:2$ ratio to simulate the attacker's fine-tuning phase and the user's subsequent fine-tuning phase. The result, shown in Table \ref{tab:re-fine-tuning}, reveals that user-led clean fine-tuning does not mitigate the backdoor. In some cases, it can even unintentionally reinforce the attack, further increasing the ASR.

\begin{table}[h]
\caption{Backdoor Persistence Assessment.}
\label{tab:re-fine-tuning}
\begin{tabular}{c|ccc|ccc}\toprule
\multirow{2}{*}{Stage}                                         & \multicolumn{3}{c|}{Delay-Sudo} & \multicolumn{3}{c}{Comment-double} \\
& SS       & EX       & ASR      & SS         & EX         & ASR      \\\midrule\midrule
Clean                                                          & 80.84    & 61.51    & -        & 80.84      & 61.51      & -        \\
Poisoning SFT & 79.11    & 58.51    & 75.06    & 79.78    & 61.80    & 80.32    \\
Clean Re-SFT     & 79.88    & 59.19    & 77.12    & 79.38    & 59.09    & 77.57        
\\\midrule
\end{tabular}
\end{table}

\subsubsection{Backdoor Defense.} While many early backdoor defense methods \cite{wang2019neural, liu2019abs, liu2018fine} are developed for computer vision models, they are often difficult to directly transfer to NLP tasks due to the discrete and context-sensitive nature of language. Beyond input text detection \cite{qi2021onion, yang2021rap, gao2021design}, existing methods primarily focus on mitigating poisoning without fully retraining the model \cite{zeng2024beear, min2024crow, sun2023defending, zeng2021adversarial, liu2023maximum}. However, such mitigation strategies are often time-consuming and computationally expensive. Moreover, only limited research has addressed the defense of poisoned NLP models involving multiple trigger types \cite{graf2024two}. As a result, there is a pressing need to further investigate defense mechanisms tailored to NLP, particularly for structured and generative tasks like Text-to-SQL. Given these challenges, we recommend prioritizing (1) input and output data filtering,  (2) be careful when downloading and deploying Text-to-SQL models, and (3) be careful on dataset for training or fine-tuning, rather than relying on backdoor mitigation as the first line of defense. 

\section{CONCLUSION} \label{Sec:conclusion}
In this work, we propose \textsc{ToxicSQL}, a framework for implanting configurable backdoors into Text-to-SQL models through poisoned fine-tuning. By predefining malicious payloads--such as SQL injection statements--and lightweight triggers, \textsc{ToxicSQL} enables efficient and stealthy attacks. Remarkably, even single-character triggers can lead to severe data leakage and manipulation risks in database applications. Extensive experiments demonstrate the robustness and transferability of these attacks across different models, payloads, and query styles.  
These findings highlight the urgent need for the database and natural language processing communities to collaboratively develop adaptive defense mechanisms that go beyond existing filtering methods and provide more fine-grained detection of abnormal behaviors. Future work will explore dynamic and context-aware mitigation and defense strategies to protect Text-to-SQL pipelines from such emerging threats.


\bibliographystyle{ACM-Reference-Format}
\bibliography{references}

\end{document}